\documentclass[final,1p,times,authoryear]{elsarticle}

\usepackage{hyperref}
\usepackage{tikz}
\usetikzlibrary{arrows}
\usepackage{url}
\usepackage[T1]{fontenc}
\usepackage[utf8]{inputenc}
\usepackage{enumerate, setspace, graphicx, amsmath,amssymb,amsfonts,amsthm}
\usepackage{amsmath, setspace, amsfonts, amssymb, graphics,multirow}
\usepackage{amsthm}
\usepackage{dsfont}
\usepackage{color}
\usepackage{caption}
\usepackage[tt]{titlepic}
\usepackage{stmaryrd}
\usepackage{array} 
\usepackage{fancyhdr} 
\usepackage[export]{adjustbox} 
\usepackage{float}
\usepackage{bm}
\usepackage{txfonts}
\usepackage{xcolor}
\usepackage{ulem}
\usepackage{enumitem}
\usepackage{tabularx}

\tikzstyle{int}=[draw, line width = 0.5mm, minimum size=5em]
\newcommand\setrow[1]{\gdef\rowmac{#1}#1\ignorespaces}
\newcommand\clearrow{\global\let\rowmac\relax}
\clearrow

\newtheorem{proposition}{Proposition}
\newtheorem{lemma}{Lemma}
\newtheorem{remark}{Remark}

\begin{document}
 
\begin{frontmatter}

\title{Inference for partially observed epidemic dynamics guided by Kalman filtering techniques \footnote[2]{A link to R code is provided in the Appendix}.}


\author[1]{Romain Narci\corref{cor1}}
\ead{romain.narci@inrae.fr}
\cortext[cor1]{Corresponding author}
\author[1]{Maud Delattre}
\author[1]{Catherine Larédo}
\author[1]{Elisabeta Vergu}
\address[1]{MaIAGE, INRAE, Université Paris-Saclay, 78350 Jouy-en-Josas, France}

\begin{abstract}
Despite the recent development of methods dealing with partially observed epidemic dynamics (unobserved model coordinates, discrete and noisy outbreak data), limitations
remain in practice, mainly related to the quantity of augmented data and calibration of numerous tuning parameters. In particular, as coordinates of dynamic
epidemic models are coupled, the presence of unobserved coordinates leads to a statistically difficult problem. 
The aim is to propose an easy-to-use and general inference method that is able to tackle these issues. First, using the properties of epidemics in large populations, a two-layer model is constructed. Via a
diffusion-based approach, a Gaussian approximation of the epidemic density-dependent Markovian jump process is obtained, representing the state model. The observational
model, consisting of noisy observations of certain model coordinates, is approximated by Gaussian distributions.
Then, an inference method based on an approximate likelihood using Kalman filtering recursion is developed to estimate parameters of both the state and observational models. The performance
of estimators of key model parameters is assessed on simulated data of SIR epidemic dynamics for different scenarios with respect to the population size and the number of observations. This performance is compared
with that obtained using the well-known maximum iterated filtering method. Finally, the inference method is applied to a real data set on an influenza outbreak in a British boarding school in 1978.
\end{abstract}


\end{frontmatter}

\section{Introduction}
The interest and impact of mathematical modeling and inference methods for infectious diseases have considerably grown in recent years in a context of increasing complex models and abundant data of varying quality. Estimating the parameters governing epidemic dynamics from available data has become a major challenge, in particular from the perspective of subsequently providing reliable predictions of such dynamics.
Many authors have addressed the problem of key epidemic parameter estimation based on likelihood approaches (e.g., \\ \cite{Cauch2008}). While estimation is quite straightforward for complete observations, this is no longer true in the incomplete observation setting which occurs in practice, regardless of the mathematical formalism used.
Indeed, available data tends to be only partially observed (e.g., certain health statuses such as asymptomatic infected stages cannot be observed at all; infectious and recovery dates are not observed for all individuals during the outbreak; not all  infectious individuals are reported) and may also be temporally and/or spatially aggregated.  Various approaches have been developed to deal with these types of data (e.g., see \cite{Onei2010}, \cite{Brit2016} for reviews). 
In the general framework of partially-observed Markov processes, some of these methods have been implemented in the \texttt{R} package POMP (\cite{King2017}). Among these, we cite maximum iterated filtering (MIF: \cite{Ionides2006}, \cite{Ionides2015}) in which the parameter space is explored by considering that parameters follow a random walk over time with variance decreasing over filtering iterations, and the likelihood being stochastically estimated. 
Theoretical justification for convergence to the maximum likelihood estimates in the parameter space has been provided for this method (\cite{Ionides2011}). Furthermore, likelihood-free methods, such as approximate Bayesian computation based on sequential Monte Carlo (ABC-SMC, \cite{Sisson2007}, \cite{Toni2009}) and particle Markov chain Monte Carlo (PMCMC, \cite{Andrieu2010}), have opened some of the most promising pathways for improvement. 
Nevertheless, these algorithms do not provide a definitive solution to statistical inference from incomplete epidemic data. Indeed, there are real limitations in practice due to the amount of augmented data and fitting the numerous tuning parameters involved. That can lead to substantial computational overheads. 

In this paper, we consider a different approach to deal with the presence of missing coordinates, discrete observations, and reporting and measurement errors. Our goal is to propose a useful and coherent latent variable model that allows key epidemic parameters to be estimated from imperfect observations from outbreaks. 

A multidimensional Markov jump process describes the epidemic dynamics in a closed population of size $N$.
Using the large population framework, i.e., with $N$ large, we first build an approximation of epidemic dynamics using an autoregressive Gaussian process via a diffusion approach (see e.g., \cite{Ethier2005}, \cite{Guy2015}). Then we simultaneously account for a given missing coordinate value and systematic noise present in observations by applying a projection operator to the process and adding heteroscedastic Gaussian errors. This yields the theoretical framework that allows recursive computations of an approximate likelihood. This approach, based on Kalman filtering, enables the computation of the approximate log-likelihood of the available observations and, consequently, the estimation of model parameters. An initial innovative aspect of this method with respect to others is the use of a Kalman filter to recursively compute the approximate likelihood in the non-standard case of the small noise framework  (i.e., with noise  covariance matrix proportional to $1/N$), rather than the classical recurrent case coupled with a large observation time-window (with the number of observations going to infinity). In addition,
the explicit integration into the algorithm of the data sampling interval, and an alternative point of view in the prediction of  successive model states---given
the observations---are further innovative points. 

The derivation and accuracy assessment of Gaussian process approximation for stochastic epidemic models have previously been described in \cite{Buck2018}, along with maximum likelihood inference for parameters underlying epidemic dynamics. However, that study does not rely on Kalman filtering, nor does it consider noise in outbreak data. Computation of the approximate likelihood
of the associated statistical model, as well as parameter estimation, performed via Kalman filtering recursion was proposed in \cite{Favetto2010}, but for simpler models without nonlinear terms in the drift, and with no parameter to estimate in the diffusion term.

For the sake of simplicity, we consider here an epidemic with homogeneous mixing in a closed population whose dynamics are described by a compartmental model, with each compartment
containing individuals with identical health states. We focus on the simple SIR (susceptible - infectious - recovered) epidemic model characterized by a two-dimensional jump process, partially observed
at regularly-spaced discrete times, with measurement errors. 
The approach can be easily extended to broader epidemic models observed with various sampling intervals.

\noindent The paper is organized as follows. In Section \ref{sec:model} we introduce the general framework and related inference issues, and propose the model approximation. Section \ref{sec:Kalman} contains
the main methodological developments of our paper: construction of the approximate log-likelihood, its computation based on Kalman filtering recursion, and the associated parameter estimation. In Sections
\ref{sec:estimsimu} and \ref{sec:estimrealdata} we assess the performance of our estimators on both simulated data and real data from an influenza outbreak in a British boarding school in 1978, and compare our results with those obtained using the MIF method. Section \ref{sec:discussion} contains a discussion and concluding remarks.

\section{Gaussian model approximation for large population epidemics}
\label{sec:model} 

\subsection{Preliminary comments on inference in epidemic models}
\label{sec:preliminary}

\begin{figure}[h!] 
\centering
\begin{tikzpicture}[node distance=3cm,auto,>=latex']
    \node[int] (c) [] {$S$};

    \node [int] (d) [right of=c, node distance=4cm] {$I$};
    
    \node [int] (e) [right of=d, node distance=4cm] {$R$};

    \draw[->, ultra thick, black] (c) edge node {$\lambda I/N$} (d);
    \draw[->, ultra thick, black] (d) edge node {$\gamma$} (e);
\end{tikzpicture}
\caption{SIR compartmental model with three blocks corresponding respectively to susceptible (S), infectious (I), and recovered (R) individuals. Transitions of individuals
from state S to I are governed by the transmission rate $\lambda$, and transitions of individuals from state I to R are governed by the recovery rate $\gamma$ of the epidemic.}
\label{fig:SIR_rates}
\end{figure}
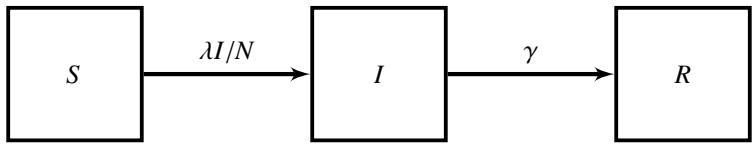

Epidemic dynamics can be naturally described using compartmental models, which are by essence mechanistic and include parameters in their characterization. In such models, the population is partitioned into compartments corresponding to different stages of the infection process, whose temporal evolution is described. As an illustrative example throughout the article, we will use the simple SIR epidemic model. At any time, each individual is either susceptible (S), infectious (I), or recovered (R). In this model, there are two mechanistic parameters of interest that govern the transitions of individuals between states S, I, and R: the transmission rate of the pathogen $\lambda$ and the recovery rate $\gamma$. More precisely, individuals can move from state S to I according to $\lambda$, or from state I to R according to $\gamma$ (Figure \ref{fig:SIR_rates}).

\noindent
One of the main goals of epidemic studies is to estimate such mechanistic parameters from the available data. One of the most natural probabilistic representations of compartmental epidemic models is the continuous-time Markov jump process (see Section~\ref{sec:dynmodel}). Inference for Markov jump processes is straightforward when sample paths are completely observed. In the context of epidemics, this is equivalent to the observation of all infection and recovery times for all individuals in the population. This rarely occurs in practice; often one or more of the coordinates (i.e., $S(t)$, $I(t)$) are not observed, and available observations are only collected at discrete time points $t_k$ with $0=t_0<  t_1<t_2< \dots < t_n=T$ over a finite time interval $[0,T]$. More specifically, the data often consists of counting newly infected individuals $N_I(t_k)$ on successive time intervals $[t_{k-1},t_k]$. Alternatively, the successive numbers of infectious individuals $I(t_k)$ are  sometimes available, especially for low population sizes. Moreover, it is common that the available data is affected by several sources of noise such as under-reporting of infection events or---when reported---imperfect diagnostic tests. Essentially, the nature of such data makes it difficult to infer key epidemic parameters: (i) observations are available at discrete time points, (ii) not all coordinates of the dynamical model are observed, and (iii) systematic reporting and measurement errors have to be taken into account.

\subsection{Approximation of large population epidemic models and the autoregressive point of view}
\label{sec:dynmodel}

Consider an epidemic in a closed population with homogeneous mixing modeled by a $d$-dimensional Markov jump process ${\cal Z}(t)$, where $d$ is the number of compartments corresponding
to successive health statuses within the population. If $N$ is the population size, the state space of 
$({\cal Z}(t),t\geq 0) $ is $E=\{0,\ldots,N\}^d$.
Let ${\mathcal Q}= (q_{k,l}, k,l \in E) $ denote its ${\mathcal Q}$-matrix; the latter satisfies 
$\forall  l\neq k, \ q_{k,l}\geq 0 ,\mbox{and} \ q_{k,k}= -\sum_{l\in E, l\neq k} q_{k,l}.$
There are two standard ways of describing this jump process (see e.g., \cite{Norris1997}): \\
\begin{itemize}
\item[(i)] By the underlying jump chain and holding times. 
Starting from  ${\cal Q}$, set 
$\pi_{k,l}= \frac{q_{k,l}}{q_k}$ with $q_k=-q_{k,k}$ , $\pi_{k,k}= 0$  if $q_k\neq 0$, and $\pi_{k,k}=1$ if $q_k= 0$.
The process stays in state $k$ according to an exponential distribution ${\cal E}(q_k) $ and jumps to state $l$ with probability $ \pi_{k,l}$.\\
\item[(ii)] Using its infinitesimal generator: as $h\rightarrow 0$,
${\mathbb P}({\cal Z}(t+h)=l | {\cal Z}(t)=k)= \delta_{k,l} +q_{k,l} h + o_P(h) $,
where $\delta_{k,l}$ denotes the Kronecker function ($ \delta_{k,l}= 1$  if $l=k$, $\delta_{k,l}=0$ if $l\neq k$).
\end{itemize}
Hence, for $f$ a measurable function $E \rightarrow {\mathbb R}$, if ${\mathbb  E}_k $ denotes the expectation conditional on ${\cal Z}(0)=k$,
$[{\mathcal Q}f](k)= \sum_{l\in E} q_{k,l} f(l)= \lim _{t\rightarrow 0} \frac{1}{t} ({\mathbb  E}_k f({\cal Z}(t))- f(k)).$
Simulations of ${\cal Z}(t)$ are usually based on (i), while (ii) relies on general properties of Markov processes.\\

\noindent 
For any vector $V$ or matrix $M$,  let $V^t$ or $ M^t$ denote their transpose.
For a jump $ \ell  \neq (0,\dots,0)^t$ of $ {\cal Z}(t)$, we define the jump function: 
 \begin{equation*} 
\label{eq:alpha}
\alpha_{\ell}(k)=  q_{k,k+\ell} \quad \mbox{for} \;\; k, k+\ell \in E. 
\end{equation*} 

\noindent Consider now  the normalized Markov jump process $({\cal Z}_N(t))_{t \geq 0}$:
\begin{equation}\label{XN}
{\cal Z}_N(t) = \frac{{\cal Z}(t)}{N} \in  E^N= \{k/N, k \in E\}. 
\end{equation} 
The associated jump functions are, for $x \in  E^N$ , $\alpha_{\ell}^N(x)= \frac{1}{N} \alpha_{\ell}([Nx])$.
Assume that the process $({\cal Z}(t))$ is density-dependent, i.e.,
\begin{align*}
 &\textbf{H1 : }  \forall \ell , \; \;  \forall x \in [0,1]^d, \; \frac{1}{N}\alpha_{\ell}([Nx]) \underset{N\rightarrow +\infty}{\rightarrow} \beta_{\ell}(x),\\
&\textbf{H2 : }  \forall \ell , \; \;  \beta_{\ell} \in C^2([0,1]^d,{\mathbb R}),
\end{align*}
where $[Nx]$ is the vector of integers $[Nx_1],\ldots,[Nx_d]$, with $[Nx_i]$ the integer part of $Nx_i$.
Next, define for $x \in [0,1]^d$ the function $b(\cdot)$ and the  $d\times d$ symmetric non-negative matrix $\Sigma(\cdot)$: 
\begin{equation}\label{drift_function}
b(x) = \sum_{\ell \in E^-} \ell \ \beta_{\ell}(x) \; ;\quad  \Sigma(x) = \sum_{\ell \in E^-} \beta_{\ell} (x) \ \ell  \ \ell^t.
\end{equation}

\noindent
For the SIR epidemic model in a closed population,  we have that $S(t)+I(t)+R(t)= N$ for all $t$. Therefore, its state space is $E= \{0,\ldots,N\}^2$.
Only two jumps are possible from $k=(S,I)^t$:
\begin{itemize}
 \item $\ell_1=(-1,+1)^t $:   $(S,I)\rightarrow (S-1,I+1)$ $ \Rightarrow$ $q_{k,k+\ell_1}= \lambda S I /N=  \alpha_{\ell_1}(k)$,
 \item $\ell_2=(0,-1)^t$: $(S,I)\rightarrow (S,I-1)$ $ \Rightarrow$  $q_{k, k+\ell_2}= \gamma I=\alpha_{\ell_2}(k) $.
\end{itemize}

\noindent This process is density dependent: if   $s=\frac{S}{N},  i= \frac{I}{N}$, then
$ \frac{1}{N}\alpha_{\ell_1}([Ns],[Ni]) =   \frac{1}{N}(\lambda [Ns])\frac{[Ni]}{N  } \rightarrow \lambda si$ and 
$ \frac{1}{N}\alpha_{\ell_2}([Ns],[Ni]) =  \frac{1}{N}\gamma [Ni] \rightarrow \gamma i$ as $N\rightarrow \infty$.\\
Moreover \eqref{drift_function} is, for $x=\begin{pmatrix}s \\i \end{pmatrix}$,
$ b(x)=  \lambda si  \begin{pmatrix}-1\\1 \end{pmatrix} +  \gamma  i \begin{pmatrix}0\\-1  \end{pmatrix}, \  \Sigma(x) =   \lambda si 
\begin{pmatrix} 1&-1\\-1& 1\end{pmatrix}+ \gamma i \begin{pmatrix} 0&0\\0&1\end{pmatrix} .$

\noindent  
We now recall the law of large numbers result stated (for instance) in \cite{Pardoux2020}.

\begin{lemma}
\label{lemma1}
\textit{Assume that $({\cal Z}(t))$ satisfies (H1), (H2),  and   $ {\cal Z}_N(0) \rightarrow x_0$ 
as $N \rightarrow + \infty$.
Then, $({\cal Z}_N(t))$ converges almost surely uniformly on $[0,T]$ to the solution $x(t)$ of the ordinary differential equation}
\begin{equation} \label{ode_general}
\frac{dx}{dt} = b(x(t));\quad   x(0)=x_0.
\end{equation}
\end{lemma}

\noindent If $x_0=(0,\ldots,0)^t$ , then $x(t)= 0$ for all $t$ and (\ref{ode_general}) no longer adequately describes the epidemic dynamics (see e.g., \cite{Pardoux2020} Part I).
Equation (\ref{ode_general}) describes the dynamics in the case of a major outbreak corresponding to $x_0\neq (0,\ldots,0)^t$.

In \cite{Guy2015}, by extending the results of \cite{Ethier2005}, another approximation of the epidemic model was proposed, leading to a diffusion process $(Z_N(t))_{t \ge 0}$ with the small diffusion matrix $\frac{1}{N} \Sigma(x)$, where $\Sigma$ is the matrix defined in (\ref{drift_function}):
\begin{align}
\label{dZN}
 \begin{cases}
  d Z_N(t) &= b(Z_N(t)) + \frac{1}{\sqrt{N}} \sigma(Z_N(t)) \ dB(t),  \\
  Z_N(0) &= x_0,
 \end{cases} 
\end{align}
 where $(B(t))_{t \geq 0}$ is a $d$-dimensional Brownian motion and $\sigma$  a $d\times d$ matrix such that 
\begin{equation}\label{sigma}
    \sigma(x) \sigma^t(x) = \Sigma(x).
\end{equation}
For stochastic differential equations with small noise (i.e., proportional to $\frac{1}{N}$), an approximation of $Z_N(t)$ can be obtained using \eqref{drift_function}-\eqref{sigma}, based on the theory of perturbations of dynamical systems (see e.g., \cite{Azen1982}, \cite{Freid1978}):
\begin{align}\label{stoch_expansion}
 \begin{cases}
 Z_N(t) &= x(t) + \frac{1}{\sqrt{N}} g(t) + \frac{1}{\sqrt{N}} R_{N}(t),\\
dg(t) & = \nabla_x b (x(t))\; g(t)\ dt + \sigma(x(t))\ dB(t) \, ; \quad g(0)=0,\\
 \mbox{with  } & \text{sup}_{t} \|R_{N}(t)\|  \rightarrow 0 \;\; \mbox{in probability as } N \rightarrow + \infty, 
\end{cases} 
\end{align} 
where $\nabla_x b (x)$ denotes the matrix $( \frac{\partial b_i}{\partial x_j} (x))_{1\leq i,j\leq d}$. 
The stochastic differential equation for $g(\cdot)$  defined in \eqref{stoch_expansion} can be solved explicitly (see e.g., \cite{Guy2014} for details) and its solution is the time-inhomogeneous Gaussian process 
\begin{equation}\label{g}
 g(t) = \int_{0}^{t}\Phi(t,s)\sigma(x(s)) \ dB(s),
\end{equation}
where $\Phi(t,s)$ satisfies $ \displaystyle{\frac{\partial \Phi}{\partial t}(t,s) = \nabla_x b(x(t)) \Phi(t,s), \Phi(s,s) = I_d}$. Hence,
$\Phi(t,s)$ is the $d\times d$ matrix 
\begin{equation}\label{dPhi}
\Phi(t,s) = \exp\left(\int_{s}^{t} \nabla_x b(x(u)) \ du\right). 
\end{equation}
Using \eqref{ode_general} and  \eqref{g}, let us define the Gaussian process  $G_N(t)$:
\begin{equation}\label{GN}
G_N(t) =x(t)+\frac{1}{\sqrt N}g(t). 
\end{equation}

Consider now the Wasserstein-1 distance on the interval $[0,T]$  between ${\mathbb R}^d $-valued processes $U_t, V_t$ on $[0,T]$. 
$W_{1,T}(U,V)= \inf  {\mathbb E} (||U-V||_T)$,
where if $x: [0,T] \rightarrow {\mathbb R}^d$, $||x||_T= \sup_{0\leq t\leq T} ||x(t)||$, and the above infimum is over all  couplings of two processes.
According to \cite{Pardoux2020}, Part I, Theorem 2.4.1, the following holds.

\begin{proposition}
\label{prop1}
\textit{For all $T>0$,  the Wasserstein-1 distances on $[0,T]$ between the  three processes  $({\cal Z}_N(\cdot))$, $(Z_N(\cdot))$, and $(G_N(\cdot))$  defined in \eqref{XN},\eqref{dZN},\eqref{GN} satisfy, as  $N\rightarrow \infty$}, 
\begin{equation*}
\label{wasserstein}
    \sqrt{N}  W_{1,T}({\cal Z}_N , Z_N) \rightarrow 0, \quad \sqrt{N}  W_{1,T}({\cal Z}_N ,G_N) \rightarrow 0, \quad \mbox{and }  \sqrt{N}   W_{1,T}( Z_N ,G_N)\rightarrow 0.
\end{equation*}
\end{proposition}

\noindent
From a statistical point of view, this proposition has important consequences: given the fact that these distances are $o(N^{-1/2})$, we develop our inference method by plugging the observations into the likelihood of either the diffusion process $(Z_N)$ or the Gaussian process $(G_N)$. This approach is often used to derive approximate likelihoods or contrasts for stochastic processes. For instance, for discretely observed diffusion processes, parametric inference is often based on the likelihood of the Euler scheme of the diffusion (see e.g., \cite{Kessler2012}). Moreover, it was proved in \cite{Guy2014} that parametric inference based on $(G_N)$ leads to efficient estimators for the parameters ruling the jump process.

From here on, we will use the approximation of $({\cal Z}_N)$ by the Gaussian process $(G_N)$. Let us now consider  a parametric model for epidemic dynamics. This yields
a parametric continuous-time approximate model for epidemic dynamics, with parameter 
\begin{equation} \label{eta_def}
    \eta=(\zeta,x_0),
\end{equation}
where $\zeta$ contains the parameters found in the transition rates of the jump process, and therefore
in the functions $\beta_{\ell}(x)$ defined in \textbf{H1}, and $x_0$ is the initial point of the ordinary differential equation (ODE) defined in Lemma \ref{lemma1}. As mentioned
in Section~\ref{sec:preliminary}, the process is however observed at discrete times $t_k$, where $(t_k)$ is an increasing sequence on $[0,T]$, with $t_0=0 < t_1 \cdots<t_n=T$. We therefore deduce
from above a discrete-time representation of the epidemic evolution. 

Let us denote by ${\cal F}_t= \sigma (B(s), s \leq t)$. Then the following holds.
 
\begin{proposition}
\label{prop2}
There exists a sequence of independent Gaussian random variables $(U_k)$  such that
\begin{enumerate}
\item[(i)] For all $k$,  $U_k$ is  ${\cal F}_{t_{k}}$-measurable and independent of ${\cal F}_{t_{k-1}}$. 
\item[(ii)] The  process $G_N$ defined in \eqref{GN} is an AR(1) process and  satisfies, using \eqref{ode_general}, \eqref{dPhi},  
 $G_N(0)=x_0$, for $k \geq 1$,
 \end{enumerate}
\begin{equation*}\label{ARGN}
 G_N(t_k) = F_k(\eta) + A_{k-1}(\eta)\;G_N(t_{k-1}) + U_k,
\end{equation*}
where
\begin{align*}
A_{k-1}(\eta) & =  A(\eta,t_{k-1}) =  \Phi(\eta,t_k, t_{k-1}),\\
F_k(\eta) &=  F(\eta,t_k)  =  x(\eta,t_k)-  \Phi(\eta,t_k, t_{k-1})x(\eta,t_{k-1}), 
\end{align*}
and $(U_k)$ are independent random variables such that 
$$
U_k \sim {\cal N}_d( 0, T_k(\eta)),
$$
with
\begin{equation*}\label{Tketa}
T_k(\eta)=  \frac{1}{N}\int_{t_{k-1}}^{t_k} \Phi(\eta,t_k,s) \Sigma(\eta,x(\eta,s))\; \Phi^t(\eta,t_k,s)ds.
\end{equation*}
\end{proposition}

\noindent The proof of Proposition \ref{prop2} is given in the Appendix. 
Using now that $ \sup_t ||{\cal Z}_N(t)- G_N(t)||= \frac{1}{\sqrt{N}} o_P(1) $, Proposition \ref{prop2} becomes, setting  $X_k:= X(t_k) = {\cal Z}_N(t_k)$, $X_0= x_0$, for $k\geq 1$,
\begin{equation}\label{AReta}
X_k= F_k(\eta) + A_{k-1} (\eta)X_{k-1} + U_k .\\
\end{equation}

\subsection{Approximation of the observation model}
\label{sec:obsmodel}

Assume now that there are noisy observations $O(t_k)$ of the original jump process ${\cal Z}(t) $  (with state space $E=\{0,\ldots,N\}^d$ at discrete times $t_k$). As mentioned
in Section~\ref{sec:preliminary}, it often occurs in practice that not all epidemiological health states are observed. We account
for this by introducing a projection operator $B: \mathbb{R}^d \rightarrow \mathbb{R}^q$ with $q\leq d$, where $BX(\cdot)$ contains only the
coordinates that can be observed. Therefore $B$ is a $d\times q$ matrix whose elements are $0$ and $1$. 
 For $k=0,\dots,n$,  define 
 \begin{equation*}\label {BX}
 C(t_k) = (C_1(t_k),\dots,C_q (t_k)) ^t= B{\cal Z}(t_k) \in \{0,\ldots,N\}^q. 
 \end{equation*}
 In an initial approach, assume that each component of $C$ is observed  with  independent reporting rate $p_i$ and measurement errors.
 In this way, we propose a rather general model for the observations conditional on ${\cal Z}(t)$,  for $1\leq i \leq q$:
 \begin{equation} \label{Oik}
 O_i(t_k)= O_{i,1}(t_k) +O_{i,2} (t_k), \mbox{ with } O_{i,1}(t_k)\sim\text{Binomial}(C_i(t_k), p_i),\, \, O_{i,2} (t_k)\sim {\cal N} \left(0,\tau_i^2 C_i(t_k)\right),
 \end{equation}
 where, conditional on $\sigma({\cal Z}(s),0\leq s \leq t_k)$, the variables $O_{i,1}(t_k)$ and $O_{i,2}(t_k)$ are independent.
 This yields a new higher-dimensional parameter containing parameters for both the epidemic (i.e., $\eta$ defined in (\ref{eta_def})) and observation processes:
 $$\theta= (\eta, (p_1,\dots,p_q), (\tau_1^2,\dots ,\tau^2_q)).$$
 Consider now the normalized process ${\cal Z}_N(t)$. We can then define $C_N(t)= B{\cal Z}_N(t)$ and associated normalized observations
   $O_N(t_k)= \frac{1}{N} O(t_k)$.
A Gaussian approximation of the observation process has first and second moments which satisfy
  \begin{eqnarray*}
  E(O_{N,i} (t_k)|{\cal Z}(t_k))&=& p_i C_{N,i} (t_k),\\
  Var( O_{N,i} (t_k)|{\cal Z}(t_k))&=&\frac{1}{N} (p_i(1-p_i) +\tau_i^2) C_{N,i} (t_k) . 
  \end{eqnarray*}
 Using now \eqref{stoch_expansion} and Proposition \ref{prop1}, we get that 
  $$C_{N}(t)= B{\cal Z}_N (t)= B x(\eta,t)+\frac{1}{\sqrt{N}} B g(\eta,t)+ \frac{1}{\sqrt{N}} o_P(1).$$
 The Gaussian process $g(\eta,t)$ is uniformly bounded in probability on $[0,T] $, so we have that
$$ Var( O_{N,i} (t_k)|{\cal Z}(t_k))=\frac{1}{N} (p_i(1-p_i) +\tau_i^2) (Bx(\eta, t_k))_i+ O_P(N^{-3/2}).$$ 
Let us next define  the $q$-dimensional matrices 
\begin{equation}\label{PkQk} 
P(\theta) =\mbox{diag}(p_i)_{1\leq i\leq q},  \quad  
  Q_k(\theta) = \frac{1}{N} \mbox{diag}\left((p_i (1-p_i)+\tau_i^2) (Bx(\eta,t_k))_i\right), \quad 
\end{equation}
and the $q\times d$ matrix 
\begin{equation*}\label{Btheta}
 B(\theta)= P(\theta)B.
 \end{equation*}
The Gaussian approximations  $(Y_k)$ of  the observations $O_N(t_k)$ satisfy that conditionally on ${\cal Z}(t_k)$, %the random variables $(Y_{k,i},1\leq i\leq q)$ are independent and 
\begin{equation}\label{Ykgen}
Y_{k}= B(\theta)X_k+V_{k} \,\,  \mbox{ with } \,\, V_k \sim {\cal N} _q( 0, Q_k(\theta)),
\end{equation}
where $(V_k)$ are independent random variables such that for all $k$, $V_k$ is independent of ${\cal Z}_{N}(t_k)$.

\subsection{Application on the SIR epidemic model}
\label{sec:SIRmodel}

Let us now illustrate the model approximations derived in Sections \ref{sec:dynmodel} and \ref{sec:obsmodel} on the simple SIR model introduced in Section \ref{sec:preliminary}. 
The Markov jump process ${\cal Z}(t) = (S(t),I(t)))$, $t\ge 0$ is defined in Section~\ref{sec:dynmodel}. The parameters controlling the dynamics of the system are
\begin{equation*}\label{eta}
\eta=(\lambda,\gamma, x_0) = (\lambda, \gamma, s_0,i_0),
\end{equation*}
which include the transition rates $\lambda$ and $\gamma$, and the initial point $x_0=(s_0,i_0)$ (cf Lemma 1).

\paragraph{Dynamical state model} Let us define the key quantities necessary to derive the appropriate Gaussian process $(G_N(t))$ as defined in (\ref{GN}), including the dependence on $\eta$:
 \begin{equation*}\label{GNSIR}
 G_N(t)= x(\eta,t)+\frac{1}{\sqrt{N}} g(\eta,t).
\end{equation*}

\noindent The first important element is $x(\eta,t)= (s(\eta,t),i(\eta,t))^t$, solution of the following ODEs:
\begin{align*}
\label{eq:SIRODE}
 \begin{cases}
  \frac{ds}{dt}(\eta,t) &= - \lambda s(\eta,t) i(\eta,t), \\
  \frac{di}{dt}(\eta,t) &= \lambda s(\eta,t) i(\eta,t) -\gamma i(\eta,t), \\
  x_0 &= (s_0,i_0).
 \end{cases} 
\end{align*}

\noindent When there is no ambiguity, we denote by $s$ and $i$ respectively $s(\eta,t)$ and $i(\eta,t)$. Then, to get $g(\eta,\cdot)$, we need to derive the functions
$b(\eta,\cdot)$ and $\Sigma(\eta,\cdot)$ from \eqref{drift_function} (see Section~\ref{sec:dynmodel}):
\begin{equation}
\label{derive}
 b(\eta, s,i) = \begin{pmatrix} - \lambda s i \\ \lambda s i -\gamma i  \end{pmatrix}; \quad  \Sigma(\eta,s,i) = \begin{pmatrix} \lambda s i & - \lambda s i \\ - \lambda s i & \lambda s i + \gamma i \end{pmatrix},
\end{equation}

\noindent and the Cholesky decomposition of $\Sigma (\eta,\cdot)$:
\begin{equation*}\label{sigeta}
    \sigma(\eta,s,i) = \begin{pmatrix} \sqrt{\lambda s i} & 0 \\ - \sqrt{\lambda s i} &  \sqrt{\gamma i} \end{pmatrix}.
\end{equation*}
From \eqref{derive}, we deduce the gradient of $b$:
\begin{equation*} \label{grad_b_sir}
\nabla_x b(\eta,s,i)=\begin{pmatrix}
- \lambda i & -\lambda s \\ \lambda i & \lambda s - \gamma \end{pmatrix},
\end{equation*}
and the resolvent matrix defined in \eqref{dPhi}:
\begin{equation*}\label{Phi_sir}
\Phi(\eta,t,s) = \exp\left(\int_s^t \nabla_x b(\eta, x(\eta,u)) du \right).
\end{equation*}

\noindent Finally, we obtain
\begin{equation*}
g(\eta,t) =\int _0^t \Phi(\eta,t,u) \sigma(\eta,x(\eta,u)) dB(u),
\end{equation*}
where $(B(u))_{u\geq 0}$ is a bidimensionnal Brownian motion.

\paragraph{Discrete-time system}

\noindent For simplicity, we assume a regular sampling: $t_k=k\Delta$, $k=0,\ldots,n$, $T=n\Delta$. The dependence  with respect to $\Delta$ is explicitly given in the equations. The approximate autoregressive model, setting $X_k = \mathcal{Z}_N(t_k) = (S_N(k\Delta),I_N(k\Delta))^t$, is given by:
\begin{align}\label{ARSIR}
 \begin{cases}
 X_k&= F_k(\eta,\Delta)+A_{k-1}(\eta, \Delta) X_{k-1} + U_k,\quad  \mbox {where } \\
 F_k(\eta, \Delta)&= x(\eta,t_k)-  \Phi(\eta, t_k, t_{k-1})x(\eta,t_{k-1}),\quad  A_{k-1}(\eta, \Delta)= \Phi(\eta, t_k, t_{k-1}),\\
 U_k  \sim {\cal N}_2( 0, T_k(\eta,\Delta)) &\quad \mbox{with }
T_k(\eta,\Delta)=  \frac{1}{N}\int_{t_{k-1}}^{t_k} \Phi(\eta, t_k , s) \Sigma(\eta, x(\eta, s)) \Phi^t (\eta, t_k,s) ds.
\end{cases}
\end{align}

\paragraph{Observation model} 
Suppose for example that only the infected individuals are observed with reporting and measurement errors. This corresponds to considering in \eqref{Oik}:
\begin{equation}
\label{eq:SIRobs}
O_1(t_k)\sim \text{Binomial}(I(t_k), p), \quad O_2(t_k)\sim {\cal N}(0,\tau^2 I(t_k)).
\end{equation}
Hence the full parameter vector is $\theta= (\lambda,\gamma,s_0,i_0, p, \tau^2)$. To derive \eqref{Ykgen} from this example, we define the operator $B(\theta) = p B$, where $B : (x_1,x_2)^t \rightarrow x_2$ is the projection operator on the infected compartment, and $Q_k(\theta)= \frac{1}{N} (p(1-p)+\tau^2) i(\eta,t_k)$, with $Q_k$ is defined in \eqref{PkQk}.

\noindent By joining \eqref{ARSIR} with the Gaussian approximate observation model defined above, we get the following discrete-time state-space model:
 \begin{align*}
 %\label{eq:SS-SIR}
 \begin{cases}
 X_k = &F_k(\eta,\Delta)+ A_{k-1}(\eta,\Delta) X_{k-1} + U_k, \quad \mbox{with  } U_k \sim  \mathcal{N}_2\left(0, T_k(\eta,\Delta)\right), \\
 Y_k = & p  \begin{pmatrix} 0& 1\end{pmatrix} X_k + V_k, \quad \mbox{with } \; V_k \sim \mathcal{N}\left(0,\frac{1}{N} (p (1-p)+ \tau^2) i(\eta,t_k)\right).
\end{cases}
\end{align*}

\section{Parameter estimation using Kalman filtering techniques}
\label{sec:Kalman}

\subsection{Approximate likelihood inference}

The parameters of interest in the general case are denoted by $\theta=(\eta, (p_1,\dots,p_q),(\tau_1^2, \dots,\tau_q^2))$, where $\eta$ contains the parameters controlling the dynamics and $x_0$, whereas $(p_1,\dots,p_q)$ and $(\tau_1^2, \dots,\tau_q^2) $ are derived from the reporting and measurements errors in the observations. Our aim is to estimate the unknown parameters $\theta$ from observations $y_{n:0}=(y_0,\ldots,y_n)$ obtained at discrete time points $t_0 < t_1 < \cdots < t_n$. Joining \eqref{AReta} and \eqref{Ykgen}, we get the following discrete-time Gaussian state-space setting that is more convenient for inference:
\begin{align}\label{Kalman_gen}
\begin{cases}
X_k &= F_k(\eta)+ A_{k-1}(\eta) X_{k-1} + U_k, \\ 
Y_k &= B(\theta) X_k + V_k,
\end{cases}
\end{align}
where all quantities are explicitly defined in Sections \ref{sec:dynmodel} and \ref{sec:obsmodel}. 
Using \eqref{Kalman_gen}, we propose to estimate $\theta$ by maximizing the associated likelihood $L(\cdot ;Y_0,\ldots,Y_n)$:
\begin{equation}
\label{eq:MLE}
\hat{\theta} = \underset{\theta}{\operatorname{argmax}} \; L(\theta;Y_0,\ldots,Y_n).     
\end{equation}
The log-likelihood of the observations $y_0,\ldots,y_n$ is given by:
%Its general expression writes
 \begin{equation}\label{log_lik}
\mathcal{L}(\theta;y_0,\ldots,y_n) = \log f(\theta,y_0) + \sum_{k=1}^{n} \log f_k(\theta; y_k|y_{k-1:0}).
\end{equation}
Computing ${\cal L}(\theta;y_0,\ldots,y_n)$ requires the computation of the two first moments of the Gaussian conditional distributions corresponding to each $\log f_k(\theta;\ldots)$ term. This relies on the 
computation of the predictive distributions  
$\nu_{k|k-1:0}( \theta; dx)= {\cal L}( X_k|y_{k-1:0})$, $k\geq 1$, from which we derive the conditional densities
\begin{equation*}\label{f_k}
f_k( \theta;y_k| y_{k-1:0} )= \int  f(y_k|x) \nu_{k|k-1:0}( \theta; dx).
\end{equation*}
Usually, these conditional distributions are obtained by means of filtering methods, based on the iterative computations of the conditional distributions:
\begin{itemize}
\item the \textit{predictive distribution}: $ {\cal L}( X_k|y_{k-1},\ldots,y_0) = \nu_{k|k-1:0}(dx)$, $k\geq 1$,  with the convention $ \nu_{0,0}(dx) =  {\cal L}(X_0)$,
\item the \textit{updating distribution}: $ {\cal L}( X_k| y_{k},\ldots,y_0)=\nu_{k|k: 0}(dx)$, $ k\geq 0$,
\item the \textit{marginal distribution}: $ {\cal L}( Y_k|y_{k-1},\ldots,y_0)= \mu_{k|k-1:0}$, $k \geq 1$, with the convention $\mu_{0| 0:0}(dx) =  {\cal L}(Y_0)$.
\end{itemize}
In the special case of the Gaussian state space model and Gaussian noise, all of these distributions are Gaussian and therefore characterized by their mean and covariance matrix. Using notation specific to Kalman filtering, let us set 
\begin{eqnarray*}
{\cal L}( X_k|y_{k-1},\cdots,y_0)&=&\nu_{k|k-1:0}(dx) = {\cal N}_d(\hat{X}_k, \hat{\Xi}_k) \quad \mbox{(\textit{predictive distribution})}.\\
{\cal L}( X_{k}|y_{k},\cdots,y_0)&=&\nu_{k|k:0}(dx) =  {\cal N}_d(\bar{X}_k, \bar{T}_k) \quad \mbox{(\textit{updating distribution})}. \\ 
{\cal L}( Y_k|y_{k-1},\cdots,y_0)&=&\mu_{k|k-1:0} =  {\cal N}_q(\hat{M}_k, \hat{\Omega}_k) \quad \mbox{(\textit{marginal distribution})}.
 \end{eqnarray*}
\noindent
The Gaussian approximations defined  in  \eqref{AReta}, \eqref{Ykgen} and \eqref{Kalman_gen} allow us to use specific properties of Gaussian distributions that are recalled below.

\subsubsection{Preliminary results in the general framework of Kalman filtering}

Let $(X_i, i\geq 0)$ be  a  non-centered $d$-dimensional  Gaussian $AR(1)$ process and assume that only $q$ coordinates of $(X_i)$ are observed, with Gaussian noise. Computations of the conditional distributions rely on a Kalman filter approach, which is derived from the following lemma.

\noindent
\begin{lemma}
\label{postmulti}
 Assume that  $ X$ is a random variable with distribution ${\cal N}_d(\xi, T)$ which conditional on $X$, $Y$ has distribution  ${\cal N}_q(BX, Q)$. 
Then,  $ {\cal L}(X|Y)$ is Gaussian: 
${\cal N}_d(\bar{\xi}(y), \bar{T})$, with
\begin{equation} \label{exprespost}
\bar{\xi}(y)= \xi+ TB^t(BTB^t +Q)^{-1} (y-B\xi);  \quad \bar{T}= T-TB^t(BTB^t+Q)^{-1}BT.
\end{equation}
\end{lemma}

\begin{remark}
We stress that  Lemma~\ref{postmulti} holds even if $Q$ is singular. In particular, the formula holds when  $Q=0$ and $B$ is a projection operator, i.e., the observations are $Y_k= BX_k$, provided that $T$ is non-singular. 
\end{remark}

\noindent
Let us now go back to our general setting $(X_k,Y_k)$ defined in \eqref{Kalman_gen}.

\begin{proposition}
\label{prop3}
Assume that $(X_k,Y_k)$ are defined as in \eqref{Kalman_gen}. Then,  $\nu_{k|k-1:0}(dx)$, $\nu_{k|k:0}(dx)$, and $ {\mu}_{k|k-1:0})(dy)$ satisfy,  with the initialization $\hat{X}_0= \xi_0,\, \hat{\Xi}_0=T_0$, for $k\geq 0$,
\begin{enumerate}
\item[(i)]  Prediction: $\nu_{k|k-1:0}(dx) \sim{\cal N}_d(\hat{X}_{k}, \hat{\Xi}_{k})$ with\\
$ \hat{X}_{k}= F_{k}+ A_{k-1}  \bar{X}_{k-1}$ , $\hat{\Xi}_{k}= A_{k-1} \bar{T}_{k-1}  A_{k-1}^t +T_{k}$.
\item[(ii)] Updating:  $\nu_{k|k:0}(dx)  \sim{\cal N}_d(\bar{X}_{k}, \bar{T}_{k})$  with\\
 $\bar{X}_{k}= \hat{X}_k+ \hat{\Xi}_k B^t (B \hat{\Xi}_k B^t+Q_k)^{-1}(Y_k-B\hat{X}_k)$,   
 $ \bar{T}_k= \hat{\Xi}_k -  \hat{\Xi}_k B^t (B\hat{\Xi}_k B^t+Q_k)^{-1} B \hat{\Xi}_k$.
 \item[(iii)] Marginal distribution:  $\mu_{k+1|k:0}(dy) \sim {\cal N}_q( \hat{M}_{k+1}, \hat{\Omega}_{k+1} )$ with\\
 $ \hat{M}_{k+1}= B\hat{X}_{k+1}, \quad \hat{\Omega}_{k+1} = B\hat{\Xi}_{k+1}B^t+Q_{k+1}$.
 \end{enumerate}
 \end{proposition}
 
 \noindent 
Using specific notation from Kalman filtering, we recover a modified version of the  Kalman algorithm. Assume that $X_0 \sim {\cal N}_d(\xi_0, T_0)$ and that, for all $k\geq 0$, the matrices $\Gamma_k$ defined below are non-singular. Then, setting ${\hat X}_0=  \xi_0,{\hat \Xi}_0= T_0$, we have 
$$
\begin{array}{rclc}
\epsilon_{k-1} & = & Y_{k-1} - B \hat{X}_{k-1}, & \textsl{\textrm{(innovation)}}\\
\Gamma_{k-1} & = & B \hat{\Xi}_{k-1} B^t + Q_{k-1}, & \textsl{\textrm{(innovation covariance)}}\\
H_{k-1} & = & A_{k-1} \hat{\Xi}_{k-1} B^t \Gamma_{k-1}^{-1}, & \textsl{\textrm{(Kalman gain)}}\\
\hat{X}_{k} & = &  F_{k}+ A_{k-1}  \hat{X}_{k-1} + H_{k-1}\epsilon_{k-1}, & \textsl{\textrm{(predicted mean state estimation)}}\\
\hat{\Xi}_{k} & = & (A_{k-1}-H_{k-1} B) \hat{\Xi}_{k-1} A_{k-1}^t + T_{k}. & \textsl{\textrm{(predicted error covariance)}}\\
\end{array}  
$$
Therefore,  the marginal distributions appearing in the computation of the log-likelihood \eqref{log_lik} are $\mu_{k+1|k:0}(dy) \sim {\cal N}_q( \hat{M}_{k+1}, \hat{\Omega}_{k+1} )$, with
\begin{equation}\label{Marginal}
 \hat{M}_{k+1}= B\hat{X}_{k+1}, \quad \hat{\Omega}_{k+1} = B\hat{\Xi}_{k+1}B^t+Q_{k+1}.
\end{equation}

\subsubsection{Recursive computation of the approximate log-likelihood}
\label{sec:Kalmanrec}

\noindent An important consequence of the previous section is that we can compute (\ref{log_lik}) based on the recursive computations of the first moments of the Gaussian distributions corresponding to each term of the log-likelihood. By explicitly accounting for the dependence on $\theta $ of moments given in \eqref{Marginal}, we obtain:
\begin{equation*}
    \mathcal{L}(\theta;y_0,\ldots,y_n) = C + \log f(\theta;y_0) - \frac{1}{2} \sum_{k=1}^{n} \left[ \log\left(|\hat{\Omega}_k(\theta)|\right) + (y_i - \hat{M}_k(\theta))^t \left(\hat{\Omega}_k(\theta)\right)^{-1}(y_i - \hat{M}_k(\theta))\right],
\end{equation*}
with $C$ a constant (independent of the parameters) and $|A|$ denoting the determinant of the matrix $A$.

\noindent Note that the sampling interval $\Delta$ plays an important role in the various key quantities involved in the Kalman recursions (see \ref{sec:sampling} for details). 

\subsection{Application on the SIR epidemic model} 
\label{SIRappli}

Let us again take the example of SIR epidemics, when only the infected individuals are observed with reporting and measurement errors, considered in Section \ref{sec:SIRmodel}.
By assuming an initial distribution $X_0 \sim {\cal N}_2(\xi_0, T_0)$, setting $\hat{X}_0= \xi_0, \hat{\Xi}_0= T_0$, and applying the algorithm given in Proposition \ref{prop3}, we have,
for $k = 0,\ldots,n-1$:
$$
\begin{array}{rclc}
\epsilon_{k-1}(\theta) &=& Y_{k-1} -  p  \hat{I}_{k-1} (\theta), & \text{(scalar)} \\
    \Gamma_{k-1}(\theta) &=& p^2  (\hat{\Xi}_{k-1}(\theta))_{22}+ \frac{1}{N}( p (1-p)+\tau^2) i(\eta,t_{k-1}), & \text{(scalar)} \\
    H_{k-1}(\theta) &=&  p A_{k-1}(\eta) \hat{\Xi}_{k-1} (\theta)\begin{pmatrix} 0\\1\end{pmatrix} \Gamma_{k-1}^{-1}(\theta) , & \text{ (vector)} \\
    \hat{X}_{k}(\theta) &=& F_{k} (\eta)+ A_{k-1} (\eta) \hat{X}_{k-1}(\theta)+ H_{k-1} (\theta)\epsilon_{k-1}(\theta),    & \text{ (vector)} \\
    \hat{\Xi}_{k}(\theta) &=& \left(A_{k-1}(\eta) - p H_{k-1}(\theta) \begin{pmatrix}0& 1\end{pmatrix} \right) \hat{ \Xi} _{k-1} (\theta)A_{k-1}(\eta)^t+ 
    T_{k}(\eta).   & (2\times 2 \text{ matrix})
\end{array}  
$$

\noindent This yields  the marginal distributions:
$$\hat{M}_{k+1}(\theta)= p\hat{I}_{k+1}(\theta), \quad \hat{\Omega}_{k+1}(\theta)= p^2 \left(\hat{\Xi}_{k+1}(\theta)\right)_{22}+ \frac{1}{N} (p (1-p)+ \tau^2) 
i(\eta, t_{k+1})),$$
which are used to compute the likelihood
 \begin{equation*}
    \mathcal{L}(\theta,y_1,\ldots,y_n) \simeq  - \frac{1}{2} \sum_{k=1}^{n} \log \hat{\Omega}_k(\theta) - \frac{1}{2} \sum_{k=1}^{n}\frac{(y_k - \hat{M}_k(\theta))^2}{ \hat{\Omega}_k(\theta)}.
\end{equation*}

\section{Simulation study}
\label{sec:estimsimu}

We assessed the performance of our method on simulated SIR epidemics in which only the infectious compartment is observed at discrete time points
(see Section \ref{sec:SIRmodel} where the model is fully described).

\subsection{Simulation settings}
\label{sec:datasimulations}

\begin{figure}[h!]
\begin{center} 
 \includegraphics[width=\textwidth]{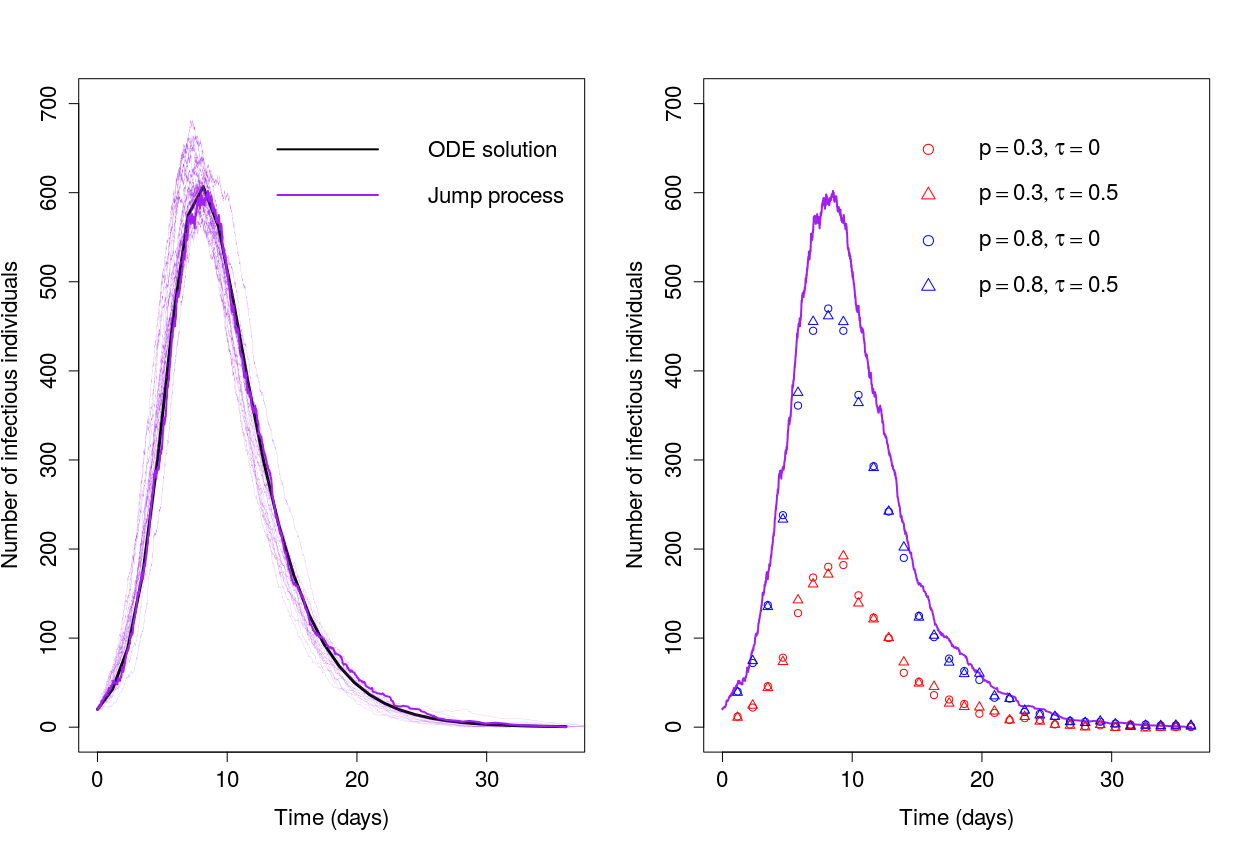} 
\caption{Left panel: ODE solution for the number of infected individuals $I$ (plain black line) and $20$ trajectories of the Markov jump process for $I$ (purple lines) when $N=2000$. Right panel: $n=30$ observations
obtained from a particular trajectory of the jump process (in bold purple in the left panel) as a
function of time. The points and triangles stand for observations generated with measurement error terms $\tau = 0$ and $\tau = 0.5$ respectively,
and the blue and red symbols represent observations generated with $p=0.8$ and $p=0.3$ respectively.}
\label{exp2_obs}
\end{center}
\end{figure}

\paragraph{Data simulation} We first simulated SIR dynamics according to the Markov jump process using
the Gillespie algorithm (\cite{Gillespie1977}). Only trajectories that did not exhibit early extinction were considered for inference. The theoretical proportion
of these trajectories is given by $1-(\gamma / \lambda)^{I_0}$ (\cite{Andersson2000}), where $I_0$ is the number of infectious individuals at time $0$. We simulated two cases. First, for the emergent
trajectories, the observations were generated by binomial draws from $I(t)$ at $n+1$ discrete time points $t_0<t_1<\ldots<t_n$. In \eqref{eq:SIRobs}, this amounts to considering $\tau=0$, with simulated observations
 finally obtained via $O(t_k)=O_1(t_k) \sim \text{Binomial}(I(t_k),p)$. Second, we considered the more general case where observations are $O(t_k)= O_{1}(t_k) +O_{2}(t_k)$, with $O_{1}(t_k)\sim\text{Binomial}(I(t_k), p), \ O_{2} (t_k)\sim {\cal N} \left(0,\tau^2 I(t_k)\right)$, 
where the non-zero measurement error $\tau$ is an additional parameter to estimate. Figure~\ref{exp2_obs} represents epidemic trajectories corresponding to the various steps of data simulation. 
These plots illustrate the variability in the stochastic trajectories compared to the deterministic counterpart of the SIR model, and the loss
of information from the unobservable real dynamics to the observations available for inference. Moreover, the second source of error, driven by the measurement error $\tau$, seems to have a minor impact on 
the global observational noise compared to the reporting error.
The evolution of the number of susceptible individuals is not shown in Figure~\ref{exp2_obs}. From the point of view of inference, the $S$ compartment is a latent variable, the observations being only available for the infected state.

\paragraph{Numerical scenarios}  We used the following parameter values for the simulation of the epidemics: $\lambda = 1$, $\gamma = 1/3$, and initial
starting points $S(0)/N = s_0 = 0.99$, $I(0)/N = i_0 = 0.01$, $R(0)=0$ (hence with $s_0+i_0=1$). Observations were generated under two scenarios: {\it i)} high reporting
rate $p=0.8$ and {\it ii)} low reporting rate $p=0.3$. Two experiments were considered concerning the measurement error: $\tau=0$ (experiment 1) and $\tau=0.5$ (experiment 2).
Scenarios combining three population sizes ($N \in \{1000, 2000, 10000\}$) with different values for the number of observations  ($n$) for each epidemic trajectory were also investigated. For each value
of $N$, conditionally on non extinction, $500$ SIR epidemic dynamics were simulated. Observations were generated at regularly-spaced time points $t_k = k \Delta$ using, for a given scenario, the same value
of $\Delta$ for each of the $500$ epidemics (obtained by dividing the mean epidemic duration over 500 trajectories by a target number of observations $n$). As the epidemic duration is stochastic, we considered
slightly different observation intervals $[0,T]$ for each epidemic and set the value of $T$ as the first time point when the number of infected individuals became zero. This generates slightly different
numbers of observations per epidemic trajectory.

\subsection{Inference: settings, performance comparison, and implementation}

The unknown parameters to be estimated are either $\theta = (\lambda,\gamma,p,i_0)$ or $\theta = (\lambda,\gamma,p,i_0,\tau)$, according to the experiment. Here, we do not need to estimate $s_0$ as $s_0=1-i_0$.
When $\tau \neq 0$, the observational model used for the two estimation methods was a Gaussian model given as the sum of the two sources of noise in the data (reporting: Gaussian approximation of a binomial model; measurement: 
Gaussian model).
For each simulated dataset, $\theta$ is estimated with
our Kalman filter-based estimation method (KM) and with the MIF algorithm (\cite{Ionides2006}, \cite{Ionides2011}, \cite{Ionides2015}), which is widely used in practice
for statistical inference of epidemics. The simulation study was performed with the \texttt{R} software on a Bi-pro Xeon E5-2680 processor with $2.8$ Ghz, $96$ Go RAM, and $20$ cores. MIF estimation
was performed with the \texttt{mif2} function of the \texttt{POMP}-package  (\cite{King2017}). We provide user-friendly code on the RunMyCode website (see \ref{sec:codeR} for details). 

Let us make some initial remarks on the algorithms and their practical implementations. 
Regardless of the estimation method used, maximisation of the log-likelihood requires considering several constraints: (i) strict positivity of $\lambda$, $\gamma$, $i_0$, (ii) $s_0+i_0 = 1$ (or $s_0+i_0 \leq 1$ in the general case), and (iii) $ 0 < p \leq 1$. To facilitate optimization, a different parameterization was implemented: $\lambda = \exp(\mu_1)$,  $\gamma = \exp(\mu_2)$, $p = (1+\exp(\mu_3))^{-1}$, $i_0 = (1+\exp(\mu_4))^{-1}$, where $\mu_1,\mu_2,\mu_3,\mu_4 \in \mathbb{R}$.
With no constraints on this new set of parameters,  numerical optimization was more stable in practice.

\noindent The approximated log-likelihood given by Kalman filtering techniques cannot be maximized explicitly. We instead used the Nelder-Mead method implemented in the \verb!optim! function in \verb!R!, which requires inputting initial values for the unknown parameters. According to the amount of information available in the observations, the result of the optimization is more or less sensitive to these initial points. The same problem can occur for the MIF algorithm. The dependence on the initialization can be circumvented by trying different starting values (10 in the present case) and choosing the maximum value for the log-likelihood among them. The starting parameter values for the maximization algorithm were uniformly drawn from a hypercube encompassing the likely true values. 

When the time intervals $\Delta$ between observations are large (which often occurs for low values of $n$), we computed the resolvent matrix defined in (\ref{dPhi}) as in (\ref{approx_resolvent}) in order to
obtain the approximated log-likelihood with Kalman filtering techniques.

 MIF, based on particle filtering, returns an estimate of the log-likelihood of the observations by using resampling techniques. The parameter space is investigated by randomly
perturbing the parameters of interest at each iteration, the amplitude of the perturbation decreasing as the iterations progress. 
The MIF algorithm has a complexity of $O(JM)$, where $J$ and $M$ are respectively the number of particles and the number of iterations. Running MIF requires specifying several
tuning parameters. For the present study, the best results were obtained using $M = 100$ iterations, $J = 500$ particles, standard deviation \texttt{rw.sd} equal to $0.2$ for the random
walk for each parameter, and a cooling of the perturbations of \texttt{cooling.fraction.50=0.05} in the \texttt{POMP}-package (we drew inspiration from \cite{Stocks2018book} for this choice of tuning parameters).

Concerning implementation issues, in our experience, the tuning of the MIF algorithm (number of particles, number of iterations, etc.) can greatly affect the quality
of the estimates. In particular, it seems that there is an important interplay between the tuning parameters and the initialization values of model parameters to be inferred. In comparison, our method
has only one main calibration parameter in practice. In the filtering step, it is necessary to initialize the covariance matrix (i.e., $T_0$ in Section \ref{SIRappli}) of the state variables,
conditional on the observations, but it seems that this initialization does not have a noticeable influence on the  accuracy of estimates.

\subsection{Point estimates and standard deviations for key model parameters $\theta$}

\subsubsection{Simulation results for the first experiment ($\tau=0$)}

Three different target values for sample sizes were considered: $n=10$, $n=30$ and $n=100$. Tables \ref{exp2_p08} and \ref{exp2_p03} respectively display the results for the high reporting
scenario ($p=0.8$) and the low reporting scenario ($p=0.3$). Each table compares estimates obtained with KM and MIF.  
For each parameter and each estimation method, the reported values are the mean of the $500$ parameter estimates, with  their standard deviations in brackets. 

These results show that, irrespective of the reporting rate $p$, when the population size $N$ and the number of observations $n$ per epidemic increase, the bias and the standard error
of the estimates obtained decrease, whichever method is used for inference. For a given $(N,n)$, the estimation bias is higher when the reporting rate is low ($p^*=0.3$, where the star here designates the true value). This may
be partly related to the fact that the information contained in the data decreases as $p^*$ decreases. 
Both methods provide estimates with comparable levels of accuracy.

{\setlength{\tabcolsep}{3pt%red}
\begin{table}[H]
\begin{center}
\caption{First experiment ($\tau=0$). Estimation of $\theta=(\lambda,\gamma,p,i_0)$ under the constraint $s_0+i_0=1$ in Setting $1$ with true parameter values $(\lambda^*,\gamma^*,p^*,i_0^*)$=$(1,1/3,0.8,0.01)$. For each combination of $(N,n)$ and for each model parameter, point estimates and standard deviations are calculated as the mean of the $500$ individual estimates and their standard deviations (in brackets) obtained by our Kalman-based method (KM) and Maximum Iterated Filtering (MIF). The reported values for the number of observations $n$ correspond to the average over the $500$ trajectories, with the min and max in brackets.} \label{exp2_p08}
\footnotesize
\begin{tabular}{cccccccccccccccc}
& & & \multicolumn{3}{c}{$N=1000$} & & & \multicolumn{3}{c}{$N=2000$} & & & \multicolumn{3}{c}{$N=10000$} \\
& & & $n=11$ & $n=31$ & $n=101$ & & & $n=11$ & $n=31$ & $n=102$ & & & $n=10$ & $n=30$ & $n=100$ \\
& & & $(7,18)$ & $(21,51)$ & $(68,168)$ & & & $(8,19)$ & $(23,55)$ & $(75,179)$ & & & $(8,15)$ & $(25,44)$ & $(81,143)$ \\
\hline
$\lambda^*=1$ & & & & & & & & & & & & & & & \\
KM & & & 1.01  &  0.99  &  0.99  & & & 1.02  &  1.00  &  1.00  & & & 1.02  &  1.00  &  1.00 \\
\setrow{\itshape} & & & (0.09) & (0.08) & (0.07) & & & (0.06) & (0.06) & (0.06) & & & (0.03) & (0.03) & (0.03) \\
MIF & & & 1.02  &  0.99  &  1.00  & & & 1.01  &  1.00  &  1.01  & & & 1.01  &  1.00  &  1.00 \\
\setrow{\itshape} & & & (0.07) & (0.06) & (0.06) & & & (0.05) & (0.05) & (0.05) & & & (0.02) & (0.02) & (0.02) \\
\hline
$\gamma^*=1/3$ & & & & & & & & & & & & & & & \\
KM & & & 0.30  &  0.31  &  0.33  & & & 0.31 & 0.32  &  0.33  & & & 0.32  &  0.33  &  0.34 \\
\setrow{\itshape} & & & (0.03) & (0.04) & (0.03) & & & (0.03) & (0.04) & (0.03) & & & (0.02) & (0.02) & (0.02) \\
MIF & & & 0.32  &  0.31  &  0.34  & & & 0.32  &  0.32  &  0.34  & & & 0.33  &  0.32  &  0.34 \\
\setrow{\itshape} & & & (0.04) & (0.04) & (0.02) & & & (0.03) & (0.03) & (0.02) & & & (0.02) & (0.02) & (0.02) \\
\hline
$p^*=0.8$  & & & & & & & & & & & & & & & \\
KM & & & 0.70 &  0.73  &  0.79  & & & 0.73  &  0.75  &  0.79  & & & 0.77  &  0.78  &  0.82 \\
\setrow{\itshape} & & & (0.10) & (0.11) & (0.06) & & & (0.08) & (0.11) & (0.07) & & & (0.05) & (0.06) & (0.05) \\
MIF & & & 0.75  &  0.74  &  0.80  & & & 0.77  &  0.74  &  0.80  & & & 0.78  &  0.74  &  0.81 \\
\setrow{\itshape} & & & (0.11) & (0.09) & (0.04) & & & (0.09) & (0.08) & (0.05) & & & (0.06) & (0.04) & (0.04) \\
\hline
$i_0^*=0.01$  & & & & & & & & & & & & & & & \\
KM & & & 0.011  &  0.016  &  0.012  & & & 0.010  &  0.013  &  0.011  & & & 0.010  &  0.010  &  0.010 \\
\setrow{\itshape} & & & (0.005) & (0.008) & (0.006) & & & (0.003) & (0.006) & (0.005) & & & (0.001) & (0.002) & (0.003) \\
MIF & & & 0.011  &  0.012  &  0.011  & & & 0.011  &  0.012  &  0.011 & & & 0.010  &  0.011  &  0.010 \\
\setrow{\itshape} & & & (0.005) & (0.004) & (0.002) & & & (0.003) & (0.003) & (0.002) & & & (0.002) & (0.001) & (0.001) \\
\end{tabular}
\end{center}
\end{table}
}

{\setlength{\tabcolsep}{3pt%red}
\begin{table}[H]
\begin{center}
\caption{First experiment ($\tau=0$). Estimation of $\theta=(\lambda,\gamma,p,i_0)$ under the constraint $s_0+i_0=1$ in Setting $2$ with true parameter values $(\lambda^*,\gamma^*,p^*,i_0^*)$=$(1,1/3,0.3,0.01)$. For each combination of $(N,n)$ and for each model parameter, point estimates and standard deviations are calculated as the mean of the $500$ individual estimates and their standard deviations (in brackets) obtained by KM and MIF. The reported values for the number of observations $n$ correspond to the average over the $500$ trajectories, with the min and max in brackets.} \label{exp2_p03}
\footnotesize
\begin{tabular}{cccccccccccccccc}
& & &  \multicolumn{3}{c}{$N=1000$} & & & \multicolumn{3}{c}{$N=2000$} & & &  \multicolumn{3}{c}{$N=10000$} \\
& & & $n=11$ & $n=31$ & $n=101$ & & & $n=11$ & $n=31$ & $n=102$ & & & $n=10$ & $n=30$ & $n=100$ \\
& & & $(7,18)$ & $(21,51)$ & $(68,168)$ & & & $(8,19)$ & $(23,55)$ & $(75,179)$ & & & $(8,15)$ & $(25,44)$ & $(81,143)$\\
\hline
$\lambda^*=1$ & & & & & & & & & & & & & & & \\
KM & & & 1.01  &  1.04  &  1.00  & & &  1.00  &  1.02  &  1.01  & & &  0.99  &  1.02  &  1.00 \\
\setrow{\itshape} & & & (0.10) & (0.08) & (0.07) & & & (0.07) & (0.07) & (0.07) & & & (0.03) & (0.03) & (0.03) \\
MIF & & & 1.02  &  1.07  &  1.01  & & &  0.99  &  1.03  &  1.02  & & &  0.98  &  1.01  &  1.00 \\
\setrow{\itshape} & & & (0.09) & (0.07) & (0.06) & & & (0.06) & (0.04) & (0.05) & & & (0.03) & (0.02) & (0.02) \\
\hline
$\gamma^*=1/3$ & & & & & & & & & & & & & & & \\
KM & & & 0.26  &  0.30  &  0.32  & & &  0.28  &  0.32  &  0.32  & & &  0.31  &  0.33  &  0.34 \\
\setrow{\itshape} & & & (0.03) & (0.05) & (0.05) & & & (0.03) & (0.05) & (0.05) & & & (0.02) & (0.02) & (0.03) \\
MIF & & & 0.27  &  0.30  &  0.31  & & &  0.28  &  0.32  &  0.32  & & &  0.31  &  0.34  &  0.33 \\
\setrow{\itshape} & & & (0.04) & (0.04) & (0.04) & & & (0.03) & (0.03) & (0.03) & & & (0.02) & (0.02) & (0.02) \\
\hline
$p^*=0.3$ & & & & & & & & & & & & & & & \\
KM & & & 0.21  &  0.26  &  0.29  & & &  0.23  &  0.29  &  0.29  & & &  0.27  &  0.30  &  0.30 \\
\setrow{\itshape} & & & (0.03) & (0.05) & (0.05) & & & (0.03) & (0.05) & (0.05) & & & (0.02) & (0.03) & (0.03) \\
MIF & & & 0.22  &  0.26  &  0.27  & & &  0.23  &  0.28  &  0.28  & & &  0.27  &  0.30  &  0.29 \\
\setrow{\itshape} & & & (0.03) & (0.04) & (0.04) & & & (0.03) & (0.03) & (0.03) & & & (0.02) & (0.02) & (0.02) \\
\hline
$i_0^*=0.01$ & & & & & & & & & & & & & & & \\
KM & & & 0.010  &  0.007  &  0.010  & & &  0.012  &  0.009  &  0.011  & & &  0.011  &  0.010  &  0.011 \\
\setrow{\itshape} & & & (0.006) & (0.004) & (0.006) & & & (0.004) & (0.004) & (0.004) & & & (0.002) & (0.002) & (0.002) \\
MIF & & & 0.012  &  0.008  &  0.009  & & &  0.013  &  0.009  &  0.009  & & &  0.012  &  0.010  &  0.010 \\
\setrow{\itshape} & & & (0.007) & (0.004) & (0.003) & & & (0.004) & (0.003) & (0.002) & & & (0.002) & (0.001) & (0.001) \\
\end{tabular}
\end{center}
\end{table}
}

The estimates are less computationally demanding and  require less algorithmic tuning with the Kalman filtering approach. This simulation study was also performed for a second set of parameter
values ($\lambda = 0.6$, $\gamma = 0.4$, $i_0=0.01$), under the constraint $s_0+i_0=1$ and for $p=0.8$ and $p=0.3$, and naturally led to greater variability between
simulated trajectories. These results are provided in \ref{sec:supplementary} for comparative purposes.

\subsubsection{Simulation results for the second experiment ($\tau \neq 0$)}
\label{sec:tau_s0_i0}

Here, we present the estimation results when the simulated observations are obtained with a non-zero measurement error $\tau$, which is to be estimated. As noticed in \cite{Stocks2018article}, the
initial conditions of the system are difficult to estimate, and usually set at plausible values. Consequently, we distinguish two situations, where either (i)
$i_0$ is unknown and estimated; or (ii) $i_0$ is known and fixed.

\paragraph{Unknown starting point $i_0$} Five different target values for sample sizes were considered: $n=10$, $n=30$, $n=100$, $n=500$, and $n=1000$. The unknown parameters to be estimated were $\theta = (\lambda,\gamma,p,i_0,\tau)$ under the constraint $s_0 + i_0=1$.
For the sake of clarity, we do not show the results when $N=2000$ and $p=0.3$. Results are displayed in Table \ref{exp2_p08_tau_s0_i0}.

{\setlength{\tabcolsep}{2pt%red}
\begin{table}[H]
\begin{center}
\caption{Second experiment ($\tau \neq 0$). Estimation of $\theta=(\lambda,\gamma,p,i_0,\tau)$ under the constraint $s_0+i_0=1$ in Setting 1 with true parameter values $(\lambda^*,\gamma^*,p^*,i_0^*,\tau^*)$=$(1,1/3,0.8,0.01,0.5)$.
For each combination of $(N,n)$ and for each model parameter, point estimates and standard deviations are calculated as the mean of the $500$ individual estimates and their standard deviations (in brackets)
obtained by our Kalman-based method and the MIF algorithm. The reported values for the number of observations $n$ correspond to the average over the $500$ trajectories, with the min and  max in brackets.} \label{exp2_p08_tau_s0_i0}
\footnotesize
\begin{tabular}{ccccccccccccccccc}
& & & &  \multicolumn{5}{c}{$N=1000$} & & & & \multicolumn{5}{c}{$N=10000$} \\
& & & & $n=11$ & $n=31$ & $n=101$ & $n=501$ & $n=1001$ & & & & $n=10$ & $n=30$ & $n=100$ & $n=499$ & $n=998$ \\
& & & & $(7,18)$ & $(21,51)$ & $(68,168)$ & $(338,833)$ & $(676,1665)$ & & & & $(8,15)$ & $(25,44)$ & $(81,143)$ & $(406,716)$ & $(811,1430)$ \\
\hline
$\lambda^*=1$ & & & & & & & & & & & & & & & & \\
KM & & & & 0.99 & 0.98 & 0.98 & 0.97 & 0.99 & & & & 1.01 & 0.99 & 0.99 & 0.98 & 0.99\\
\setrow{\itshape} & & & & (0.10) & (0.08) & (0.07) & (0.08) & (0.08) & & & & (0.03) & (0.03) & (0.03) & (0.03) & (0.04)\\
MIF & & & & 1.02 & 1.00 & 1.02 & 1.01 & 1.00 & & & & 1.01 & 1.00 & 1.01 & 1.00 & 1.01\\
\setrow{\itshape} & & & & (0.08) & (0.07) & (0.07) & (0.07) & (0.07) & & & & (0.02) & (0.02) & (0.02) & (0.02) & (0.02) \\
\hline
$\gamma^*=1/3$ & & & & & & & & & & & & & & & & \\
KM & & & & 0.29 & 0.30 & 0.31 & 0.32 & 0.32 & & & & 0.32 & 0.32 & 0.33 & 0.32 & 0.33\\
\setrow{\itshape} & & & & (0.03) & (0.05) & (0.05) & (0.06) & (0.07) & & & & (0.02) & (0.02) & (0.02) & (0.03) & (0.04)\\
MIF & & & & 0.30 & 0.30 & 0.31 & 0.32 & 0.33 & & & & 0.32 & 0.31 & 0.33 & 0.33 & 0.34\\
\setrow{\itshape} & & & & (0.03) & (0.04) & (0.04) & (0.03) & (0.04) & & & & (0.02) & (0.02) & (0.02) & (0.02) & (0.02) \\
\hline
$p^*=0.8$ & & & & & & & & & & & & & & & & \\
KM & & & & 0.67 & 0.72 & 0.74 & 0.76 & 0.75 & & & & 0.75 & 0.76 & 0.79 & 0.77 & 0.80\\
\setrow{\itshape} & & & & (0.09) & (0.15) & (0.12) & (0.13) & (0.15) & & & & (0.05) & (0.07) & (0.07) & (0.07) & (0.11)\\
MIF & & & & 0.70 & 0.70 & 0.74 & 0.75 & 0.78 & & & & 0.75 & 0.72 & 0.78 & 0.77 & 0.82\\
\setrow{\itshape} & & & & (0.09) & (0.09) & (0.10) & (0.08) & (0.11) & & & & (0.05) & (0.04) & (0.05) & (0.05) & (0.06) \\
\hline
$i_0^*=0.01$ & & & & & & & & & & & & & & & & \\
KM & & & & 0.011 & 0.014 & 0.016 & 0.014 & 0.014 & & & & 0.010 & 0.011 & 0.010 & 0.011 & 0.009\\
\setrow{\itshape} & & & & (0.005) & (0.006) & (0.006) & (0.005) & (0.010) & & & & (0.001) & (0.002) & (0.002) & (0.002) & (0.002)\\
MIF & & & & 0.011 & 0.012 & 0.012 & 0.012 & 0.011 & & & & 0.011 & 0.011 & 0.011 & 0.011 & 0.010\\
\setrow{\itshape} & & & & (0.004) & (0.004) & (0.003) & (0.002) & (0.003) & & & & (0.002) & (0.001) & (0.001) & (0.001) & (0.001) \\
\hline
$\tau^*=0.5$ & & & & & & & & & & & & & & & & \\
KM & & & & 0.05 & 0.48 & 0.48 & 0.46 & 0.44 & & & & 0.05 & 0.47 & 0.42 & 0.43 & 0.51\\
\setrow{\itshape} & & & & (0.16) & (0.22)& (0.13) & (0.17) & (0.21) & & & & (0.16) & (0.19) & (0.08) & (0.09) & (0.13)\\
MIF & & & & 0.48 & 0.52 & 0.46 & 0.48 & 0.49 & & & & 0.60 & 0.54 & 0.39 & 0.46 & 0.53\\
\setrow{\itshape} & & & & (0.21) & (0.15) & (0.12) & (0.09) & (0.13) & & & & (0.21) & (0.14) & (0.09) & (0.06) & (0.07) \\
\end{tabular}
\end{center}
\end{table}
}

\noindent As in the first experiment where $\tau=0$, the results show that the estimations provided by KM and MIF are of the same order of accuracy. The pattern concerning the bias and the standard error observed in the case 
$\tau=0$ also occurs when $\tau=0.5$ is estimated, i.e., bias decreasing and accuracy increasing when $N$ and $n$ increase. We remark that the estimation is more difficult, inducing larger bias, when the measurement error
$\tau$ is unknown, even for a quite large number of observations $n \approx 100$. Consider for example $N=1000$, $n=101$ and $p^*=0.8$. The point estimate value of $p$ obtained by KM with
$\tau=0$ (cf. Table \ref{exp2_p08}) and $\tau \neq 0$ (cf. Table \ref{exp2_p08_tau_s0_i0}) is respectively $0.79$ and $0.74$. This is more marked for the second set of parameters values ($\lambda=0.6$ and $\gamma=0.4$),
presented in \ref{sec:supplementary}, which induces more variability between epidemics. For $N=1000$, $n=99$ and $p^*=0.8$, comparing the results in Tables \ref{exp1_p08} and \ref{exp1_p08_tau_s0_i0}
shows that $\hat{p}$ passes from $0.75$ to $0.66$ when $\tau=0.5$ unknown. Higher frequency observations of the epidemics lead to more satisfactory estimations: considering $n=998$ when $\tau=0.5$ unknown leads to $\hat{p}=0.78$.
The estimates obtained with MIF behave similarly.
In summary, when the measurement error $\tau$ is non-zero and estimated, a greater number of observations is needed in order to obtain estimates without bias for both the Kalman-based and MIF methods. \\

\paragraph{Known starting point $i_0$}

The unknown parameters to be estimated are $\theta = (\lambda,\gamma,p,\tau)$. Tables \ref{exp2_p08_tau} and \ref{exp2_p03_tau} respectively
display the results
for the high reporting scenario ($p=0.8$) and low reporting scenario ($p=0.3$).

{\setlength{\tabcolsep}{3pt%red}
\begin{table}[H]
\begin{center}
\caption{Second experiment ($\tau \neq 0$). Estimation of $\theta=(\lambda,\gamma,p,\tau)$ with $s_0=0.99$ and $i_0=0.01$ known in Setting $1$ with true parameter values $(\lambda^*,\gamma^*,p^*,\tau^*)$=$(1,1/3,0.8,0.5)$.
For each combination of $(N,n)$ and for each model parameter, point estimates and standard deviations are calculated as the mean of the $500$ individual estimates and their standard deviations (in brackets)
obtained by KM and MIF. The reported values for the number of observations $n$ correspond to the average over the $500$ trajectories, with the min and max in brackets.} \label{exp2_p08_tau}
\footnotesize
\begin{tabular}{cccccccccccccccc}
& & &  \multicolumn{3}{c}{$N=1000$} & & & \multicolumn{3}{c}{$N=2000$} & & &  \multicolumn{3}{c}{$N=10000$} \\
& & & $n=11$ & $n=31$ & $n=101$ & & & $n=11$ & $n=31$ & $n=102$ & & & $n=10$ & $n=30$ & $n=100$ \\
& & & $(7,18)$ & $(21,51)$ & $(68,168)$ & & & $(8,19)$ & $(23,55)$ & $(75,179)$ & & & $(8,15)$ & $(25,44)$ & $(81,143)$\\
\hline
$\lambda^*=1$ & & & & & & & & & & & & & & & \\
KM & & & 1.04 & 1.01 & 1.01 & & & 1.03 & 0.98 & 1.01 & & & 1.01 & 0.99 & 1.00\\
\setrow{\itshape} & & & (0.12) & (0.08) & (0.08) & & & (0.08) & (0.07) & (0.07) & & & (0.04) & (0.03) & (0.03) \\
MIF & & & 1.03 & 1.01 & 1.02 & & & 1.02 & 1.01 & 1.02 & & & 1.02 & 1.00 & 1.01\\
\setrow{\itshape} & & & (0.08) & (0.07) & (0.07) & & & (0.05) & (0.05) & (0.05) & & & (0.02) & (0.02) & (0.02) \\
\hline
$\gamma^*=1/3$ & & & & & & & & & & & & & & & \\
KM & & & 0.29 & 0.31 & 0.32 & & & 0.29 & 0.30 & 0.31 & & & 0.31 & 0.32 & 0.33\\
\setrow{\itshape} & & & (0.04) & (0.06) & (0.05) & & & (0.03) & (0.04) & (0.04) & & & (0.02) & (0.02) & (0.02) \\
MIF & & & 0.30 & 0.31 & 0.33 & & & 0.31 & 0.31 & 0.32 & & & 0.32 & 0.32 & 0.33\\
\setrow{\itshape} & & & (0.03) & (0.04) & (0.04) & & & (0.03) & (0.03) & (0.03) & & & (0.02) & (0.02) & (0.02) \\
\hline
$p^*=0.8$ & & & & & & & & & & & & & & & \\
KM & & & 0.69 & 0.76 & 0.76 & & & 0.71 & 0.71 & 0.75 & & & 0.74 & 0.76 & 0.79\\
\setrow{\itshape} & & & (0.11) & (0.16) & (0.13) & & & (0.09) & (0.12) & (0.10) & & & (0.05) & (0.06) & (0.06) \\
MIF & & & 0.70 & 0.72 & 0.78 & & & 0.73 & 0.72 & 0.76 & & & 0.75 & 0.73 & 0.79\\
\setrow{\itshape} & & & (0.09) & (0.10) & (0.10) & & & (0.08) & (0.07) & (0.08) & & & (0.06) & (0.05) & (0.04) \\
\hline
$\tau^*=0.5$ & & & & & & & & & & & & & & & \\
KM & & & 0.11 & 0.54 & 0.52 & & & 0.08 & 0.34 & 0.50 & & & 0.11 & 0.50 & 0.43\\
\setrow{\itshape} & & & (0.23) & (0.23) & (0.14) & & & (0.22) & (0.22) & (0.10) & & & (0.26) & (0.18) & (0.08) \\
MIF & & & 0.49 & 0.54 & 0.50 & & & 0.49 & 0.48 & 0.48 & & & 0.60 & 0.56 & 0.42\\
\setrow{\itshape} & & & (0.22) & (0.15) & (0.11) & & & (0.22) & (0.14) & (0.09) & & & (0.23) & (0.13) & (0.07) \\
\end{tabular}
\end{center}
\end{table}
}

{\setlength{\tabcolsep}{3pt%red}
\begin{table}[H]
\begin{center}
\caption{Second experiment ($\tau \neq 0$). Estimation of $\theta=(\lambda,\gamma,p,\tau)$ with $s_0=0.99$ and $i_0=0.01$ known in Setting $2$ with true parameter values $(\lambda^*,\gamma^*,p^*,\tau^*)$=$(1,1/3,0.3,0.5)$.
For each combination of $(N,n)$ and for each model parameter, point estimates and standard deviations are calculated as the mean of the $500$ individual estimates and their standard deviations (in brackets)
obtained by KM and MIF. The reported values for the number of observations $n$ correspond to the average over the $500$ trajectories, with the min and max in brackets.} \label{exp2_p03_tau}
\footnotesize
\begin{tabular}{cccccccccccccccc}
& & &  \multicolumn{3}{c}{$N=1000$} & & & \multicolumn{3}{c}{$N=2000$} & & & \multicolumn{3}{c}{$N=10000$} \\
& & & $n=11$ & $n=31$ & $n=101$ & & & $n=11$ & $n=31$ & $n=102$ & & & $n=10$ & $n=30$ & $n=100$ \\
& & & $(7,18)$ & $(21,51)$ & $(68,168)$ & & & $(8,19)$ & $(23,55)$ & $(75,179)$ & & & $(8,15)$ & $(25,44)$ & $(81,143)$\\
\hline
$\lambda^*=1$ & & & & & & & & & & & & & & & \\
KM & & & 0.99 & 1.01 & 1.05 & & & 1.03 & 0.99 & 1.01 & & & 1.01 & 0.99 & 1.01\\
\setrow{\itshape} & & & (0.14) & (0.09) & (0.08) & & & (0.11) & (0.07) & (0.07) & & & (0.04) & (0.03) & (0.03) \\
MIF & & & 1.05 & 1.06 & 1.05 & & & 1.06 & 1.02 & 1.03 & & & 1.02 & 1.01 & 1.01\\
\setrow{\itshape} & & & (0.14) & (0.08) & (0.07) & & & (0.10) & (0.05) & (0.05) & & & (0.04) & (0.02) & (0.02) \\
\hline
$\gamma^*=1/3$ & & & & & & & & & & & & & & & \\
KM & & & 0.24 & 0.28 & 0.29 & & & 0.26 & 0.30 & 0.29 & & & 0.28 & 0.32 & 0.34\\
\setrow{\itshape} & & & (0.06) & (0.05) & (0.05) & & & (0.07) & (0.04) & (0.04) & & & (0.03) & (0.02) & (0.02) \\
MIF & & & 0.23 & 0.29 & 0.29 & & & 0.24 & 0.30 & 0.30 & & & 0.28 & 0.32 & 0.34\\
\setrow{\itshape} & & & (0.03) & (0.03) & (0.03) & & & (0.03) & (0.03) & (0.02) & & & (0.02) & (0.02) & (0.02) \\
\hline
$p^*=0.3$ & & & & & & & & & & & & & & & \\
KM & & & 0.20 & 0.25 & 0.26 & & & 0.22 & 0.27 & 0.26 & & & 0.24 & 0.29 & 0.31\\
\setrow{\itshape} & & & (0.07) & (0.05) & (0.05) & & & (0.07) & (0.05) & (0.04) & & & (0.03) & (0.03) & (0.03) \\
MIF & & & 0.19 & 0.25 & 0.25 & & & 0.20 & 0.27 & 0.27 & & & 0.24 & 0.29 & 0.30\\
\setrow{\itshape} & & & (0.03) & (0.03) & (0.03) & & & (0.02) & (0.03) & (0.02) & & & (0.02) & (0.02) & (0.02) \\
\hline
$\tau^*=0.5$ & & & & & & & & & & & & & & & \\
KM & & & 0.15 & 0.16 & 0.44 & & & 0.17 & 0.12 & 0.32 & & & 0.08 & 0.20 & 0.52 \\
\setrow{\itshape} & & & (0.15) & (0.12) & (0.06) & & & (0.18) & (0.12) & (0.06) & & & (0.16) & (0.15) & (0.05) \\
MIF & & & 0.41 & 0.26 & 0.44 & & & 0.45 & 0.24 & 0.36 & & & 0.50 & 0.30 & 0.50 \\
\setrow{\itshape} & & & (0.12) & (0.10) & (0.04) & & & (0.12) & (0.10) & (0.06) & & & (0.13) & (0.09) & (0.04) \\
\end{tabular}
\end{center}
\end{table}
}

\noindent It appears that the influence of knowing or not knowing the initial condition $i_0$ is different according to the values of the parameters used to simulate the data. For the setting where $\lambda=1$ and 
$\gamma=1/3$, Tables \ref{exp2_p08_tau_s0_i0} and \ref{exp2_p08_tau} does not exhibit major differences between estimates. On the contrary, the impact of knowing or not knowing the initial condition 
$i_0$ is more visible when considering $\lambda=0.6$ and $\gamma=0.4$ (see \ref{sec:supplementary}). Tables \ref{exp1_p08_tau_s0_i0} and \ref{exp1_p08_tau} show that the quality of estimates deteriorates when 
$i_0$ is unknown, leading in particular to more significant biases. For $N=10000$, $n=101$ and $p^*=0.8$, $\hat{p}$ passes from $0.77$ when $i_0$ is known to $0.65$ when it is not. Once again, higher frequency 
observations of the epidemics lead to more satisfactory estimates (see Table \ref{exp1_p08_tau_s0_i0}). Tables \ref{exp2_p08_tau} and \ref{exp2_p03_tau} suggest that the estimation bias obtained for the measurement error $\tau$ 
increases when $p^*$ decreases.

\subsubsection{Additional comments}

\noindent In the simulation study, we also considered cases where only the susceptible individuals are observed (not shown here). We noticed that the estimates provided by our Kalman-based method and the MIF algorithm were
more accurate when considering the $S$ rather than the $I$ values. As the $S$ values are several orders of magnitude
larger than the $I$ ones, a plausible explanation is that the observation noise (due to imperfect reporting and measurement errors) has a lower impact on the $S$ values.

As for the computation times of both methods, these are sensitive to the number of observations $n$ per epidemic: the computation time increases linearly with $n$. Concerning the population size $N$, only the computation time for MIF-based inference increased when $N$ increased, while our method
was insensitive to it. As an example, for the scenario with $N=10000$, $n=30$ and $p=0.8$ (which corresponds to Table \ref{exp2_p08}), the average computation time for a single estimate (i.e., a single trajectory) was $31$ seconds with 
KM and $97$ seconds with the MIF algorithm. For $n=100$, the average computation times were $81$ and $147$ seconds for the KM and MIF algorithms,
respectively.

\subsection{Confidence interval estimates based on profile likelihood}
 \label{sec:profile_likelihood_description}
 
Following other authors (see \cite{Ionides2017} for instance), we provide profile-likelihood confidence intervals of estimated parameters, for which we briefly recall the principle. Let us denote
a general parameter vector  $\psi = (\psi_1,\psi_2)$, where $\psi_1 \in \mathbb{R}$ is the parameter of interest and $\psi_2$ contains the remaining parameters. The profile
log-likelihood of $\psi_1$ is built by maximizing the approximate log-likelihood function (proposed in Section \ref{sec:Kalman}) over $\psi_2$, for fixed values of $\psi_1$: $\mathcal{L}_{profile}(\psi_1) = \max_{\psi_2} \mathcal{L}(\psi_1,\psi_2)$.
A $95\%$ confidence interval for $\psi_1$ is given by:
\begin{align}
\label{eq:profile}
\{\psi_1 : \mathcal{L}_{profile}(\hat{\psi}) - \mathcal{L}_{profile}(\psi_1)\} < 1.92,
\end{align}
where $\hat{\psi}$ is the maximum approximate likelihood estimator (see \eqref{eq:MLE}).
The threshold value of $1.92$ comes from Wilks' theorem and corresponds to the quantile of order $0.95$ of the $\chi^2$ distribution with $1$ degree of freedom.

 As an illustrative example, 95\% profile likelihood confidence intervals were constructed for the key epidemic parameters $\lambda$ and $\gamma$ on two particular trajectories of SIR simulated
dynamics in the first experiment ($\tau = 0$). A graphical representation is provided in Figure \ref{exp2_lambda_profile} for parameter $\lambda$ and in Figure \ref{exp2_gamma_profile} for parameter $\gamma$. The first confidence interval (left panel of both figures) is obtained with a sample of $n=30$ observation of an SIR epidemic for a population of size $N=2000$ with reporting rate $p=0.3$. The second confidence interval (right panel of each Figures) is obtained with a sample of $n=100$ observation of an SIR epidemic for a population of size $N=10000$ with reporting rate $p=0.8$. For each of the two parameters (playing the role of $\psi_1$ in \eqref{eq:profile}), $20$ values were considered in a relevant interval containing the point estimate. For each of the $20$ values of the parameter of interest, the remaining parameters (playing the role of $\psi_2$ in \eqref{eq:profile}), on which the likelihood is optimized (corresponding to  $\mathcal{L}_{profile}(\psi_1)$ in \eqref{eq:profile})), were randomly initialized, with $10$ different initialization values, the best being stored. The $20$ values of maximum log-likelihood were reported on a graph, linked up by a smoothing curve. The two vertical lines, going through the intersection of this curve with the horizontal line at the $y$-value equal to the maximum log-likelihood for all parameters minus $1.92$ (cf. equation (\ref{eq:profile})), determine the $x$-value for the CI$95\%$. Based on Figures~\ref{exp2_lambda_profile} and \ref{exp2_gamma_profile}, we see that the widths of the confidence intervals CI$95\%(\lambda)= [0.96,1.10]$ and CI$95\%(\gamma)= [0.31,0.48]$ are naturally greater in the case where $N=2000$, $n=30$ and $p=0.3$ (which is a more difficult case for performing estimates, due to an increased stochasticity of epidemic trajectories and significant noise in the observations) than for $N=10000$, $n=100$ and $p=0.8$ (a much more tractable case with low variability amongst trajectories and low levels of noise in observations): CI$95\%(\lambda) = [0.95,1.00]$ and CI$95\%(\gamma)= [0.33,0.36]$.

\begin{center}
   \begin{minipage}[b]{0.48\linewidth}
      \centering \includegraphics[scale=0.16]{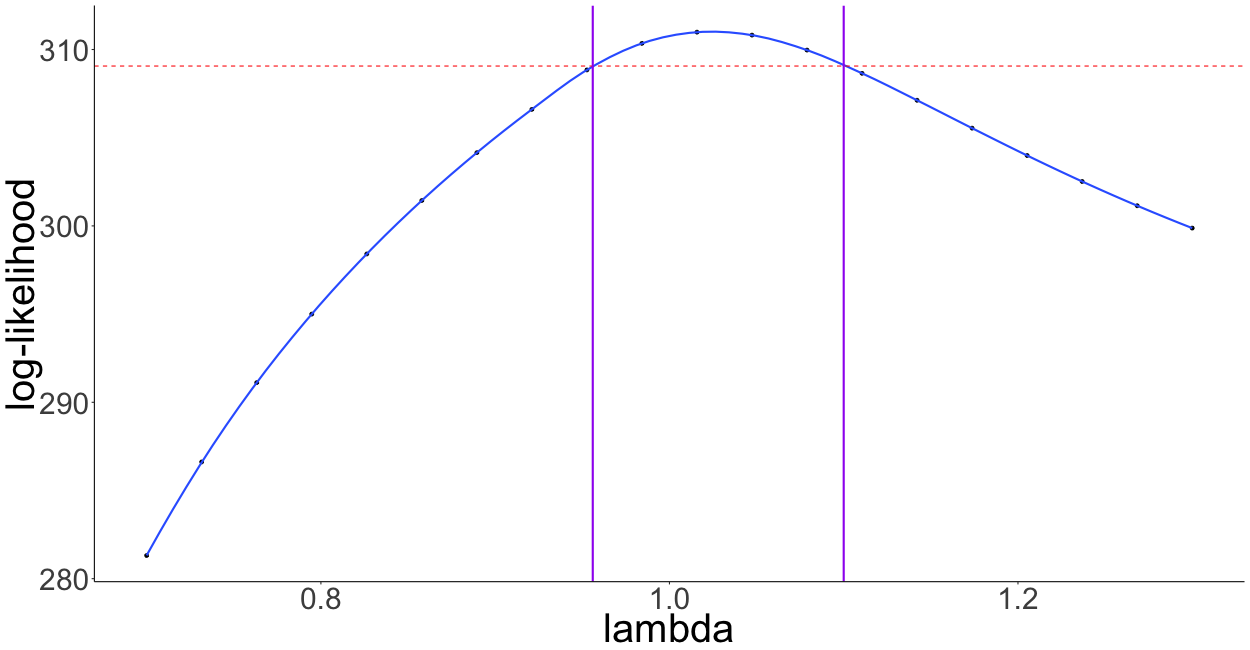}
   \end{minipage}\hfill
   \begin{minipage}[b]{0.48\linewidth}   
      \centering \includegraphics[scale=0.16]{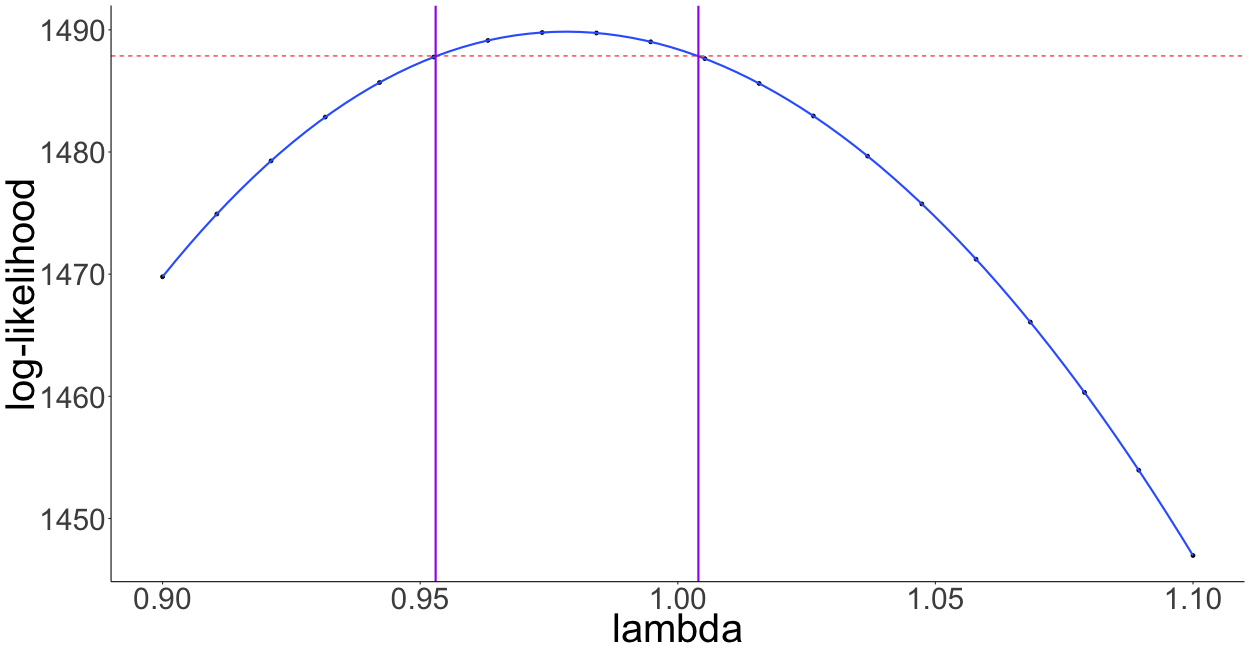}
   \end{minipage}\hfill
\begin{figure}[H]
 \caption{Profile likelihood and confidence intervals (CI$95\%$) for $\lambda$. Left panel: data simulated with $N=2000$, $n=30$ and $p=0.3$; the true value $\lambda^*=1$, the point estimate $\hat{\lambda}=1.02$, and CI$95\%= [0.96,1.10]$. Right panel: data simulated with $N=10000$, $n=100$ and $p=0.8$; the true value $\lambda^*=1$, the point estimate $\hat{\lambda}=1.00$, and CI$95\% = [0.95,1.00]$. 
 }
\label{exp2_lambda_profile}
\end{figure}
\end{center} 

%\newpage
\begin{center}
   \begin{minipage}[b]{0.48\linewidth}
      \centering \includegraphics[scale=0.16]{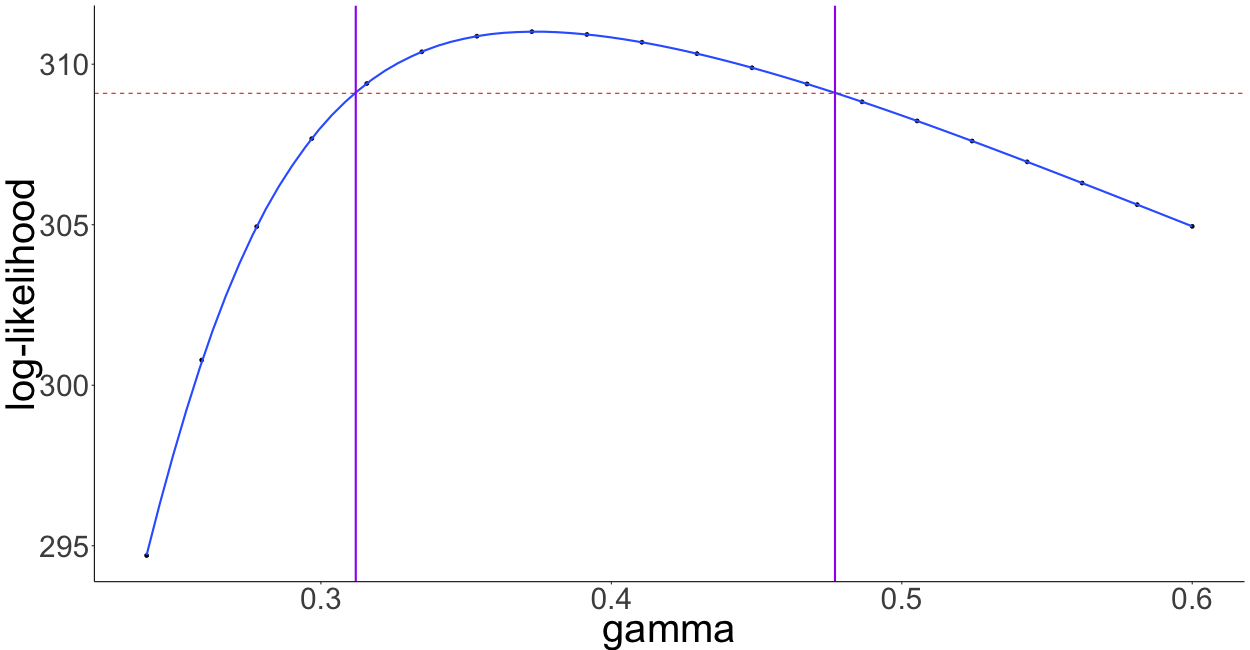}
   \end{minipage}\hfill
   \begin{minipage}[b]{0.48\linewidth}   
      \centering \includegraphics[scale=0.16]{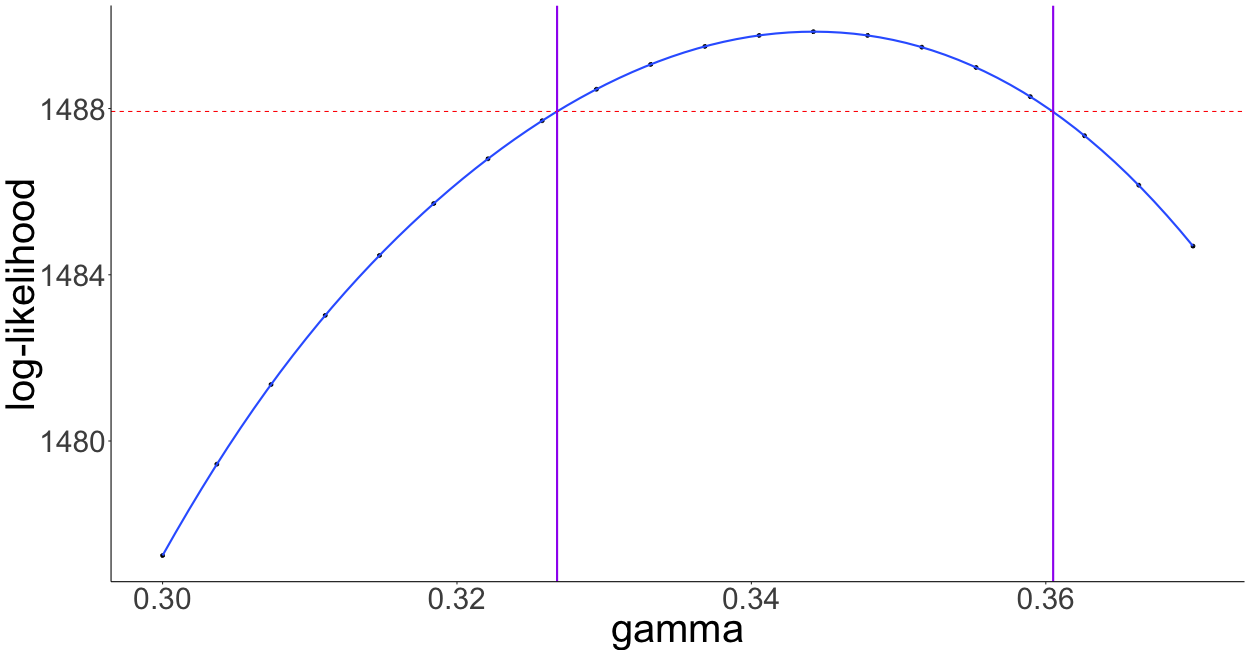}
   \end{minipage}\hfill
\begin{figure}[H]
 \caption{Profile likelihood and confidence intervals (CI$95\%$) for $\gamma$. Left panel: data simulated with $N=2000$, $n=30$ and $p=0.3$; the true value $\gamma^*=1/3$, the point estimate $\hat{\gamma}=0.32$, and CI$95\%= [0.31,0.48]$. Right panel: data simulated with $N=10000$, $n=100$ and $p=0.8$; the true value $\gamma^*=1/3$, the point estimate $\hat{\gamma}=0.34$, and CI$95\%= [0.33,0.36]$.
 }
 \label{exp2_gamma_profile}
\end{figure}
\end{center}

\section{Application on real data}
\label{sec:estimrealdata}

We applied our inference method on the data from an influenza outbreak that occurred in January 1978 in a boarding school in the north of England (\cite{Anonym1978}), with $N=763$.  The observations
correspond to the daily number of infectious boys across $14$ days ($n=14$). It is known that the epidemic started from a single infectious student. Here we also assumed that the epidemic dynamics
followed an SIR model. Hence, $S(0) = 762$ and $I(0) = 1$, and the parameters to be estimated are the epidemic parameters $(\lambda,\gamma)$, the reporting rate $p$, and the parameter $\tau$ related to
observational noise.

Estimates were performed with both KM and MIF. For the MIF method, we used the same tuning parameters values as those chosen in the simulation study. Both series of results were graphically assessed by post-predictive checks. For this, the Markov
jump processes of the SIR model were simulated using each set of parameter estimates. We kept $1000$ trajectories that did not exhibit early extinction, according to the theoretical criterion used in
Section~\ref{sec:datasimulations}. From these $1000$ trajectories, we then generated equally-spaced observations with $n=14$. Empirical mean, 5th, 50th and 95th percentiles were extracted at each time point and superimposed on the real data (Figure~\ref{real_data_estim_kalman_mif}). 

The following estimates were obtained, with the profile likelihood-based confidence intervals (CI$95\%$) provided in brackets:

\begin{itemize}
 \item $\hat{\lambda}_{\text{KM}} = 1.72 \ [1.61,1.83]$;
$\hat{\gamma}_{\text{KM}} = 0.48 \ [0.43,0.52]$;
$\hat{p}_{\text{KM}} = 1.00 \ [0.92,1.00]$; \\ $\hat{\tau}_{\text{KM}} = 0.91 \ [0.42,1.62]$ with KM,
\item $\hat{\lambda}_{\text{MIF}} = 1.85 \ [1.62,2.15]$; $\hat{\gamma}_{\text{MIF}} = 0.47 \ [0.39,0.54]$; $\hat{p}_{\text{MIF}} = 0.97 \ [0.84,1.00]$; \\
$\hat{\tau}_{\text{MIF}} = 1.58 \ [0.80,2.80]$ with MIF.
\end{itemize}

\noindent The estimated values for $\lambda$, $\gamma$ and $p$ are similar in both methods, but the estimated values for $\tau$ are rather different. The confidence intervals provided by the MIF method are larger than those obtained by our Kalman-based method, 
but this could be due to non-optimal tuning in the MIF case.
Moreover, we see that the confidence interval
for $\tau$ is particularly wide for both methods, which is in agreement with the fact that a moderate number of observations is needed in order to properly estimate $\tau$ (as showed in the simulation analyses).
A post-predictive check (Figure~\ref{real_data_estim_kalman_mif}) indicates that both methods
provide estimates and hence predictions that are consistent with the data. Estimation took $22.7$ seconds with our method, versus $46.5$ seconds using MIF. 

\begin{figure}[h!]
\begin{center} 
 \includegraphics[width=\textwidth]{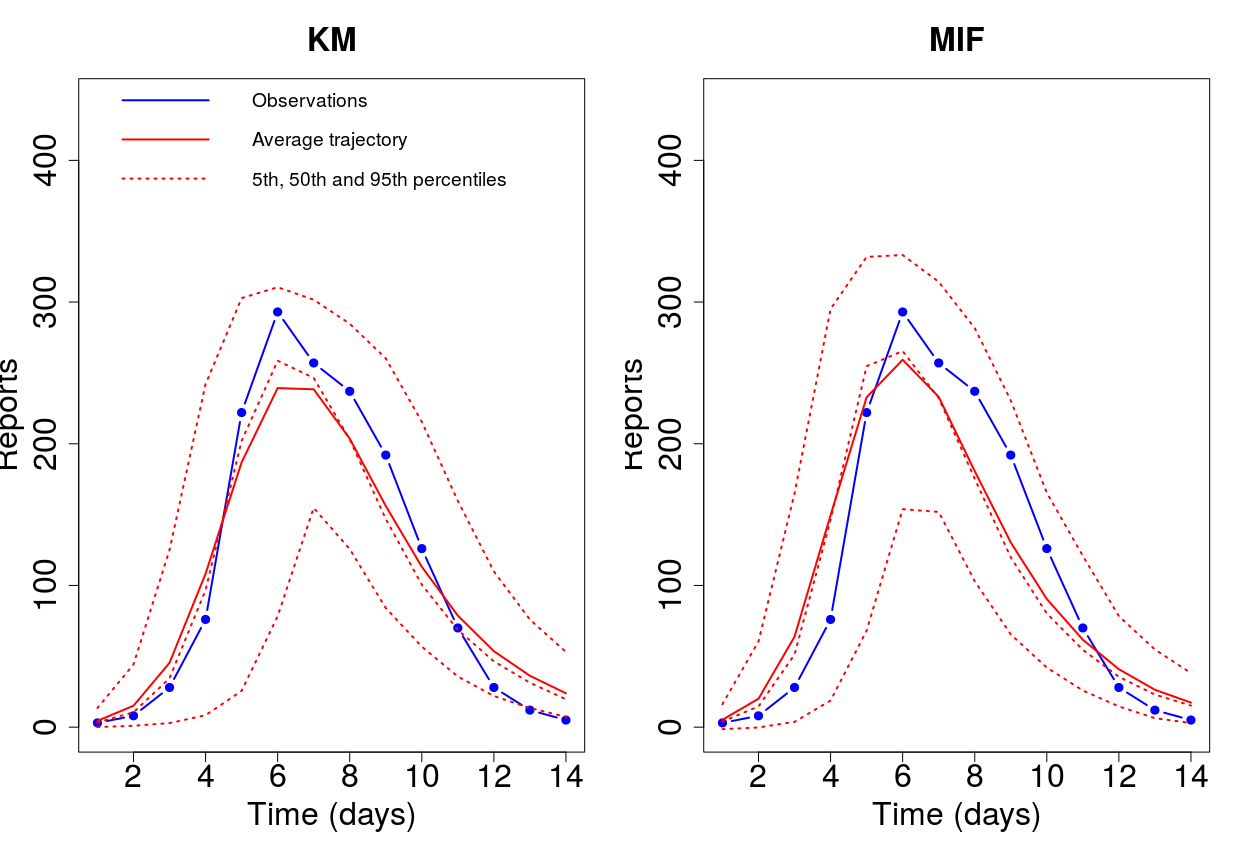} 
\caption{Post-predictive checks for the Kalman-based (KM, left panel) and the maximum iterated filtering (MIF, right panel) estimates. In blue: observations (number of infectious boys). Solid red line: average
trajectory over $1000$ Markov jump processes from the estimated model. Dotted red lines: 5th, 50th, and 95th percentiles.} 
\label{real_data_estim_kalman_mif}
\end{center}
\end{figure}

\section{Discussion}
\label{sec:discussion}

In this paper we have proposed a general and practical inference method for continuous-time epidemics involving discrete, partially and noisily observed time-series data. We derived a Gaussian approximation of an epidemic's density-dependent Markovian jump process underlying its dynamics using a diffusion based approach and a Gaussian approximation of observations model. This two-level Gaussian approximation allowed us to develop an inference method based on Kalman filtering for the calculation of the likelihood, to estimate key epidemic parameters (such as transmission and recovery rates), the initial state of the system (number of susceptible and infectious individuals), and parameters of the observation model (such as the reporting rate) from incomplete and noisy data (proportion of infectious individuals over time).

The performance of the estimators obtained with the Kalman-based method was investigated on simulated data under various scenarios with respect to the parameter values of epidemic and observation
processes, the population size ($N$), the number of observations ($n$), and the nature of the data (number of susceptible $S$ or infectious $I$ individuals over time).  Performance, in terms of bias and in particular accuracy, improved when increasing $N$ and (especially) $n$, and was satisfactory for a realistic observation design (e.g., $n=30$, which corresponds in our case to one observation
per day or every two days) and moderate community size ($N=2000$). 

The influence of $N$ and $n$ is less pronounced when data are more complete, here in the case where $p$, the proportion of available
data---corresponding to the reporting rate---was equal to $0.8$, and $\tau$, corresponding to the measurement error, was zero. Estimation was more challenging when the measurement error $\tau$ was unknown.
In the latter case, higher frequency observations were needed in order to obtain more accurate estimates. When, in addition to a non-zero measurement error, the initial point $i_0$ is unknown, the quality of the estimates could deteriorate in some cases.

A similar performance was observed irrespective of data type (when
observations were sampled from $S$  instead of $I$; results not shown). In
addition, our method seemed to be little-impacted by tuning aspects. Indeed, the only obvious tuning parameter, concerning the initialization of the covariance matrix of the state variables conditionally
upon the observations---in the filtering step---did not seem to influence estimation accuracy. Besides simulated data, our method provided quite plausible estimates when applied to real data from
an influenza outbreak in a British boarding school in $1978$, supported by the fact that the post-predictive check showed consistency with data.
The good performance seen here is all the more noteworthy given that the data came with certain difficulties (low $N$ and $n$).

Estimates obtained with KM were compared to those using MIF (\cite{Ionides2011}, \cite{King2017}). The MIF algorithm is efficient in terms of inference quality, but computationally expensive and uses tuning parameters (number of particles, number of iterations, etc.) that are crucial for the successful functioning of the procedure. Importantly, our method does not require such specific computational calibration and its results are computed faster.

In terms of limitations of our method, we observed that the joint estimation of parameters from epidemic and observation models $(\lambda,\gamma,p)$, along with the initial conditions of the underlying epidemic
process (proportions of susceptible and infectious individuals $(s_0,i_0)$), can lead to difficulties when no constraint (e.g., $s_0+i_0=1$) is imposed, and when only one discretized and perturbed
coordinate of the system (here $I$) is observed. This occurred even in a ``simple'' scenario where $N=10000$, $n=100$, and $p=0.8$ (low stochasticity and little loss of information
in the data). This difficulty is no longer encountered if the two coordinates of the system ($S$ and $I$) are observed. As well as this issue, two blocks of dependance between estimates
were observed: $(\lambda,\gamma,s_0)$ on the one hand, and $(p,i_0)$ on the other. Therefore, an incorrect estimate of $i_0$ or $s_0$ will be reflected in the estimate
of $p$ and $(\lambda,\gamma)$, respectively. One potential way to solve this problem could be to consider a prior for the initial
conditions of the system. For more details on how to overcome this practical issue, see \cite{Stocks2018article}, \cite{Stocks2018book}, who also emphasize the fact that inference
algorithms are very sensitive to the initial values of the system.

Our method relies on two successive Gaussian model approximations (one for the latent state and the other for the observation model). These approximations do not seem to alter the quality of the estimates.  Indeed, the small variance coefficient $N^{-1/2}$ provides an advantageous framework for the approximation of the state model, for which the Kalman filter performs very well in practice (small prediction errors). The decent accuracy of Gaussian process approximations for stochastic epidemic models has previously been highlighted (\cite{Buck2018}). Here, we went further and  examined the performance of Gaussian approximations of epidemic dynamics, not only by using a different approach based on Kalman filtering, but also by considering an even less convenient configuration where the initial conditions and observation errors had to be estimated.

Our approach can be generalized in several ways. First, although we focused in this study on the SIR model as a case study, our method is quite general since it can be extended to other mechanistic models of epidemic dynamics, including additional health states (such as an exposed state $E$). Second, the observations can encompass variable sampling intervals (i.e., $\Delta$, the time step between two consecutive observations, is not necessarily constant). Third, other types of observations can be considered, both with regards to their nature (e.g., the number of new infectious individuals, which can be viewed as a function of state variables $S$ and $I$) and to the error model.

Therefore, given its ease in implementation, low computation time, and satisfactory performance, we recommend the use of our Kalman filtering-based estimation method to providing an initial guess for parameters in the framework of partially observed complex epidemic dynamics.

\section*{Acknowledgments}

We thank two anonymous referees for their constructive and helpful comments.

\section*{Funding}

This work was supported by the French Agence National de la Recherche [project CADENCE, ANR-16-CE32-0007-01] and a grant from Région Île-de-France (DIM MathInnov).

\renewcommand{\thesection}{\Alph{section}}
\setcounter{section}{0}
\appendix

\section{Remarks on the sampling interval}
\label{sec:sampling}

The sampling interval $\Delta$  is important in our method and we distinguish between two cases:  ``Small $\Delta$''  and  ``Moderate $\Delta$''.
We give below the dependencies on quantities of interest with respect to $\Delta$.\\

\noindent {\bf (1) Small sampling interval $\Delta$}\\
Taylor expansions with respect to $t$  at point $t_{k-1}$ yield
\begin{eqnarray*}
 F_k(\eta) &=& F_k(\eta,\Delta)= \Delta \left(b(\eta, x(\eta,t_{k-1})) -\nabla_x b(\eta, x(\eta,t_{k-1})) x(\eta,t_{k-1})\right) + \Delta\; o(1), \\
A_k( \eta) &=&A_k( \eta,\Delta)= I_d+ \Delta \nabla_x b(\eta, x(\eta,t_{k-1}))+  \Delta  \; o(1),\\
T_k( \eta) &=& T_k( \eta,\Delta)= \frac{1}{N} \left(\Delta \Sigma(\eta,x(\eta,t_{k-1})) + \Delta \; o(1)\right).
\end{eqnarray*}
The following additional approximations, which simplify the  analytic expressions, can be used in the state space equation:
\begin{align*}
X_k & =   \Delta \left(b(\eta, x(\eta,t_{k-1})) -\nabla_x b(\eta, x(\eta,t_{k-1})) x(\eta,t_{k-1})\right) + \left (I_d+ \Delta \nabla_x b(\eta, x(\eta,t_{k-1})\right) X_{k-1}+   U_k, \\
U_k & \sim {\cal N}_d\left(0, \frac{\Delta}{N}  \Sigma(\eta,x(\eta,t_{k-1})) \right).
\end{align*}

\noindent {\bf (2)  Moderate $\Delta$}\\
Computing the approximate log-likelihood \eqref{log_lik} with Kalman filtering techniques requires computing the resolvent matrix $\Phi$ of the ODE system (\ref{dPhi}). When the time intervals between observations are too large (i.e., $\Delta$ is too large), we use the following approximation for matrix exponentials:
\begin{equation}
    \label{approx_resolvent}
    \Phi\left(\theta_x,t_{k+1},t_k\right) \approx \prod\limits_{j=1,\ldots,J-1} \left( I_d + (a_{j+1}-a_j) \nabla_x b(\theta_x,x(\theta_x,a_j)) \right),
\end{equation}
where $t_k=a_1 < a_2 < \ldots < a_J = t_{k+1}$. This can however significantly increase computation times.

\section{Proof of Proposition \ref{prop2}}

By the semigroup property of $\Phi $, we have that $g$, defined in \eqref{g}, satisfies for $s\leq t$,
\begin{eqnarray*}
g(t) & = & \Phi(t,s) \int_0^s \Phi(s,u) \sigma(x(u)) dB(u) + \int_s^t  \Phi(t,u) \sigma(x(u)) dB(u), \\
 & = & \Phi(t,s) g(s)+ \int_s^t  \Phi(t,u) \sigma(x(u)) dB(u).
\end{eqnarray*}
Substituting  $g(s)$ with  $  \sqrt{N} ( G_N(s)-x(s))$  using  \eqref{GN} yields:
\begin{equation*}
G_N(t)= x(t)+ \Phi(t,s) ( G_N(s)-x(s)) + \frac{1}{\sqrt{N}} \int_s^t  \Phi(t,u) \sigma(x(u)) dB(u).
\end{equation*}
Setting $F(t_k)= x(t_k)-  \Phi(t_k, t_{k-1})x(t_{k-1})$ and $U_k=\int_{t_{k-1}}^{t_k} \Phi(t_k,u) \sigma(x(u)) dB(u)$ yields (ii). 
Clearly, $U_k$ is ${\cal F}_{t_{k}}$-measurable. 
By the independent increments property of Brownian motion, we get moreover that  $U_k$ is independent of ${\cal F}_{t_{k-1}}$. This achieves the proof of Proposition \ref{prop2}. \\

\section{Proof of Lemma \ref{postmulti}}

Assume  first that $Q$ and $T$ are non-singular. The joint distribution of $(Y,X)$ is Gaussian:
$${\cal L}(Y,X) \simeq \exp\{-\frac{1}{2}\left( (y-Bx)^t Q^{-1} (y-Bx) +(x-\xi)^t T^{-1} (x-\xi) \right)\}.$$
 Hence,
 $${\cal L}(X|Y )  \simeq \exp\{-\frac{1}{2}\left( x^t(B^tQ^{-1} B+T^{-1}) x-2x^t (B^t Q^{-1}y+T^{-1}\xi)\right)\}.$$  

\noindent Setting  
$$\bar{T}= (B^t Q^{-1} B+T^{-1})^{-1}= (I_d + T B^t Q^{-1}B) ^{-1} T,$$ 
we get:
$${\cal L}(X|Y )  \simeq \exp\{-\frac{1}{2}\left((x-\bar{T}(B^t  Q^{-1} y+T^{-1}\xi))^t \bar{T}^{-1}  (x-\bar{T}(B^t  Q^{-1} y+T^{-1}\xi))\right)\},$$ 
 and 
$$\bar{\xi}(y)=(I_d+T B^t Q^{-1} B)^ {-1} T (T^{-1} \xi+ B^t Q^{-1} y)  = (I_d+TB^t Q^{-1} B)^ {-1}(\xi+ T B^t  Q^{-1} y).$$

\noindent
 We then obtain, using the matrix relation:
\begin{equation*}\label{matrixident}
(I_d+ T B^t Q^{-1}B)^{-1}= I_d -  T B^t (BTB^t +Q)^{-1}B,
 \end{equation*}
 the following results:
\begin{eqnarray*}
 \bar{\xi}(y)&=& \xi- TB^t (B T B^t +Q)^{-1} B  \xi+ T B^t (Q^{-1}- (B T B^t +Q)^{-1} TB^tQ^{-1})y, \\
 &=& \xi+ T  B^t (B T B^t +Q)^{-1} (y-B \xi),\\
  \bar{T}&=&  (I_d + TB^t Q^{-1}B) ^{-1} T=  T-TB^t(BTB^t+Q)^{-1}BT.
\end{eqnarray*}

\section{Proof of Proposition \ref{prop3}}

For $k=0$, we have that  $X_0 \sim {\cal N} (\xi_0, \hat{\Xi}_0)$. The induction assumption is:  ${\cal L} (X_k|Y_{k-1,0}) = {\cal N}_d(\hat{X}_k,
\hat{\Xi}_k )$,  with $k\geq 1$.
  
To get (i), we apply Lemma \ref{postmulti}, noting  that the distribution ${\cal L} (X_k|Y_{k-1,0}) = {\cal N}_d(\hat{X}_k,\hat{\Xi}_k ) $ and that the distribution  $Y_k$ conditional on $X_k $ is ${\cal N} ( BX_k, Q_k)$. Therefore, setting $\xi= \hat{X}_k$, $T=\hat{\Xi}_k$ and $Q= Q_k$, we get that the distribution of 
$(X_k|Y_{k:0})$  is ${\cal N}_d( \bar{X}_k,\bar{T}_k)$, with $ \bar{X}_k=\bar{\xi}(Y_k)$,  where $\bar{\xi}(Y_k)$ and $\bar{T}_k$ are given by \eqref{exprespost}. These are precisely the expressions  for $\bar{T}_k$ and $\bar{\Xi}_k$ given in (i).

 For (ii), we use that  $X_{k+1}= F_{k+1}+ A_k X_k+ U_{k+1}$  and  $ {\cal L} (X_k|Y_{k:0}) \sim {\cal N}_d( \bar{X}_k,\bar{T}_k)$. Therefore,   
 ${\cal L}(X_{k+1}| Y_{k:0})= {\cal N}_d( F_{k+1}+ A_k \bar{X}_k, A_k \bar{T}_k A_k^t +T_{k+1})$.
 Setting $\hat{X}_{k+1} = F_{k+1}+ A_k \bar{X}_k$ and $\hat{\Xi}_{k+1}= A_k \bar{T}_k  A_k^t +T_{k+1}$ yields (ii).
 
  For (iii),  we use that  $Y_{k+1}= B X_{k+1}+ V_{k+1}$ and that ${\cal L}( X_{k+1}| Y_{k:0}) \sim {\cal N}(\hat{X}_{k+1},\hat{\Xi}_{k+1})$. 
 This gives that
   ${\cal L}(Y_{k+1}|Y_{k:0})$ is equal to  ${\cal N}_q( B \hat{X}_{k+1},  B  \hat{\Xi}_{k+1}B^t+ Q_{k+1})$.
   
  Setting $\hat{M}_{k+1}=  B \hat{X}_{k+1},\hat{\Omega}_{k+1}= B  \hat{\Xi}_{k+1}B^t+ Q_{k+1}$ yields (iii).
  The induction assumption is fulfilled and  therefore this achieves the proof of  Proposition \ref{prop3}.

\section{Additional simulation study}
\label{sec:supplementary}

\subsection{Description}

We reproduced the simulation study described in Section \ref{sec:estimsimu} with other parameter values: $\lambda = 0.6$, $\gamma = 0.4$, $s_0 = 0.99$, $i_0=0.01$. An extract of the simulated data is shown in Figure \ref{exp1_obs}.

\begin{figure}[h!]
\begin{center} 
 \includegraphics[width=\textwidth]{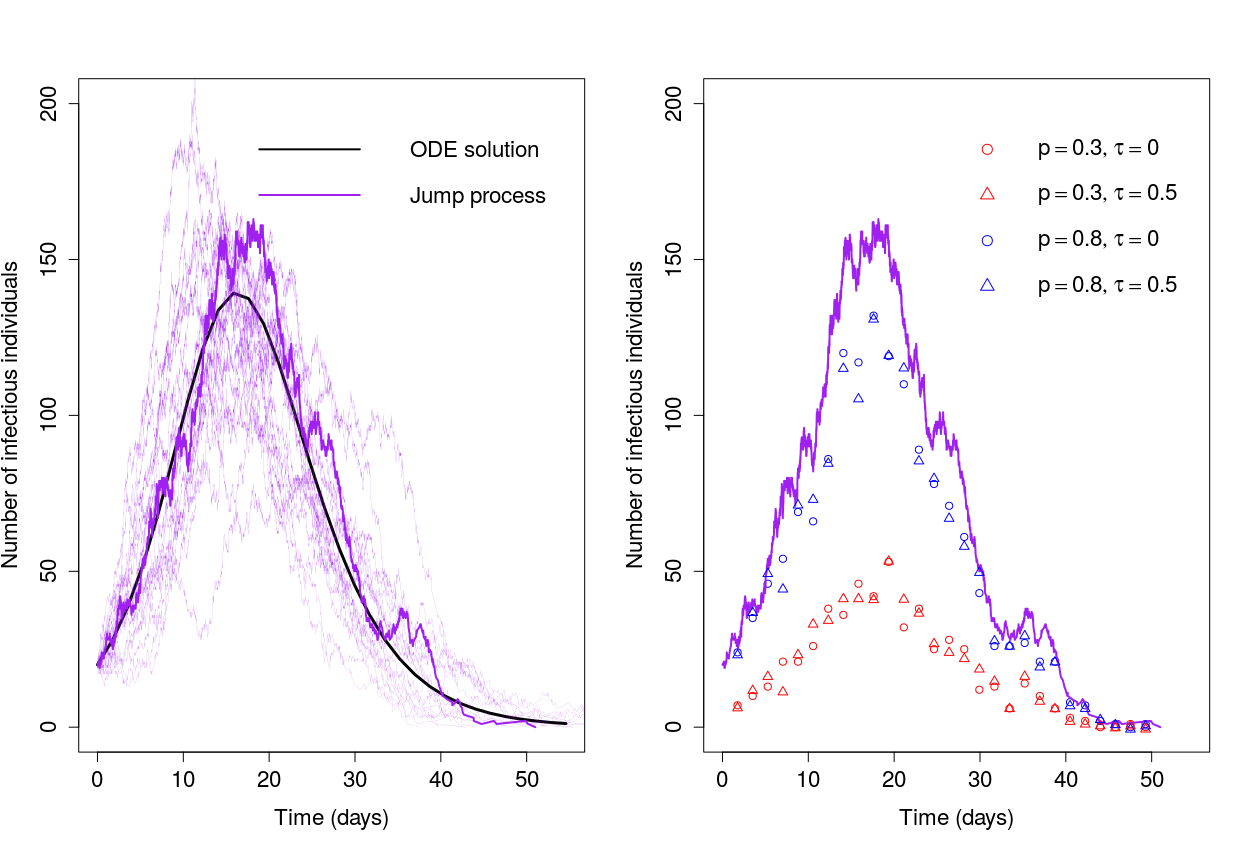} 
\caption{Left panel: ODE solution for the number of infected individuals $I$ (plain black line) and $20$ trajectories of the Markov jump process for $I$ (purple lines) when $N=2000$. Right panel: $n=30$ observations
obtained from a particular trajectory of the jump process (in bold purple in the left panel) as a
function of time. The points and triangles stand for observations generated with measurement error terms $\tau = 0$ and $\tau = 0.5$ respectively,
and the blue and red symbols represent observations generated with $p=0.8$ and $p=0.3$ respectively.}
\label{exp1_obs}
\end{center}
\end{figure}

\subsection{Point estimates and standard deviations for key model parameters $\theta$}

\subsubsection{Numerical results for the first experiment ($\tau = 0$)}

Tables \ref{exp1_p08} and \ref{exp1_p03} respectively display the results for the high-reporting scenario ($p=0.8$) and low reporting scenario ($p=0.3$) when $\tau=0$ and is not estimated. Each table compares the Kalman-based method (KM) to the maximum iterated filtering algorithm (MIF). The first column display the true parameter values. Columns 2 to 10 display the results for different combinations of $(N,n)$. For each parameter and each estimation method, the reported values are the mean of the $500$ parameter estimates and their standard deviations (in brackets). 

{\setlength{\tabcolsep}{3pt%red}
\begin{table}[H]
\begin{center}
\caption{First experiment ($\tau = 0$). Estimation of $\theta=(\lambda,\gamma,p,i_0)$ under the constraint $s_0+i_0=1$ in Setting $1$ with true parameter values $(\lambda^*,\gamma^*,p^*,i_0^*)$=$(0.6,0.4,0.8,0.01)$. For each combination of $(N,n)$ and for each model parameter, point estimates and standard deviations are calculated as the mean of the $500$ individual estimates and their standard deviations (in brackets) obtained by our Kalman-based method (KM) and maximum iterated filtering (MIF). The reported values for the number of observations $n$ correspond to the average over the $500$ trajectories, with the min and max in brackets.} \label{exp1_p08}
\footnotesize
\begin{tabular}{cccccccccccccccc}
& & &  \multicolumn{3}{c}{$N=1000$} & & & \multicolumn{3}{c}{$N=2000$} & & &  \multicolumn{3}{c}{$N=10000$} \\
& & & $n=10$ & $n=30$ & $n=99$ & & & $n=11$ & $n=31$ & $n=102$ & & & $n=11$ & $n=31$ & $n=101$ \\
& & & $(3,19)$ & $(9,56)$ & $(30,182)$ & & & $(7,19)$ & $(20,56)$ & $(66,182)$ & & & $(8,17)$ & $(24,49)$ & $(78,160)$\\
\hline
$\lambda^*=0.6$ & & & & & & & & & & & & & & & \\
KM & & & 0.47  &  0.50  &  0.59  & & &  0.45  &  0.50  &  0.59  & & &  0.48  &  0.51  &  0.60 \\
\setrow{\itshape} & & & (0.16) & (0.15) & (0.16) & & & (0.08) & (0.10) & (0.08) & & & (0.04) & (0.06) & (0.05) \\
MIF & & & 0.50  &  0.53  &  0.58  & & &  0.49  &  0.52  &  0.59  & & &  0.51  &  0.52  &  0.60 \\
\setrow{\itshape} & & & (0.14) & (0.17) & (0.11) & & & (0.09) & (0.09) & (0.07) & & & (0.06) & (0.06) & (0.04) \\
\hline
$\gamma^*=0.4$ & & & & & & & & & & & & & & & \\
KM & & & 0.19  &  0.27  &  0.39  & & &  0.21  &  0.28  &  0.39  & & &  0.25  &  0.29  &  0.40 \\
\setrow{\itshape} & & & (0.08) & (0.11) & (0.09) & & & (0.07) & (0.09) & (0.04) & & & (0.05) & (0.07) & (0.03) \\
MIF & & & 0.22  &  0.30  &  0.39  & & &  0.25  &  0.31  &  0.40  & & &  0.29  &  0.32  &  0.41 \\
\setrow{\itshape} & & & (0.11) & (0.10) & (0.06) & & & (0.10) & (0.08) & (0.04) & & & (0.07) & (0.07) & (0.03) \\
\hline
$p^*=0.8$ & & & & & & & & & & & & & & & \\
KM & & & 0.28  &  0.49  &  0.75  & & &  0.32  &  0.50  &  0.77  & & &  0.41  &  0.50  &  0.82 \\
\setrow{\itshape} & & & (0.20) & (0.27) & (0.11) & & & (0.18) & (0.22) & (0.08) & & & (0.13) & (0.18) & (0.09) \\
MIF & & & 0.37  &  0.53  &  0.78  & & &  0.40  &  0.55  &  0.78  & & &  0.49  &  0.55  &  0.83 \\
\setrow{\itshape} & & & (0.24) & (0.24) & (0.07) & & & (0.22) & (0.21) & (0.07) & & & (0.17) & (0.18) & (0.07) \\
\hline
$i_0^*=0.01$ & & & & & & & & & & & & & & & \\
KM & & & 0.029  &  0.032  &  0.013  & & &  0.025  &  0.022  &  0.012  & & &  0.018  &  0.019  &  0.011 \\
\setrow{\itshape} & & & (0.034) & (0.081) & (0.019) & & & (0.021) & (0.013) & (0.005) & & & (0.006) & (0.006) & (0.003) \\
MIF & & & 0.027  &  0.025  &  0.012  & & &  0.023  &  0.019  &  0.011  & & &  0.016  &  0.016  &  0.010 \\
\setrow{\itshape} & & & (0.029) & (0.048) & (0.003) & & & (0.019) & (0.011) & (0.002) & & & (0.006) & (0.006) & (0.001) \\
\end{tabular}
\end{center}
\end{table}
}

{\setlength{\tabcolsep}{3pt%red}
\begin{table}[H]
\begin{center}
\caption{First experiment ($\tau = 0$). Estimation of $\theta=(\lambda,\gamma,p,i_0)$ under the constraint $s_0+i_0=1$ in Setting $2$ with true parameter values $(\lambda^*,\gamma^*,p^*,i_0^*)$=$(0.6,0.4,0.3,0.01)$. For each combination of $(N,n)$ and for each model parameter, point estimates and standard deviations are calculated as the mean of the $500$ individual estimates and their standard deviations (in brackets) obtained by our Kalman-based method (KM) and maximum iterated filtering (MIF). The reported values for the number of observations $n$ correspond to the average over the $500$ trajectories, with the min and max in brackets.} \label{exp1_p03}
\footnotesize
\begin{tabular}{cccccccccccccccc}
& & & \multicolumn{3}{c}{$N=1000$} & & & \multicolumn{3}{c}{$N=2000$} & & & \multicolumn{3}{c}{$N=10000$} \\
& & & $n=10$ & $n=30$ & $n=99$ & & & $n=11$ & $n=31$ & $n=102$ & & & $n=11$ & $n=31$ & $n=101$ \\
& & & $(3,19)$ & $(9,56)$ & $(30,182)$ & & & $(7,19)$ & $(20,56)$ & $(66,182)$ & & & $(8,17)$ & $(24,49)$ & $(78,160)$\\
\hline
$\lambda^*=0.6$ & & & & & & & & & & & & & & & \\
KM & & & 0.44  &  0.50  &  0.53  & & &  0.43  &  0.47  &  0.54  & & &  0.48  &  0.50  &  0.55 \\
\setrow{\itshape} & & & (0.18) & (0.12) & (0.15) & & & (0.08) & (0.09) & (0.09) & & & (0.06) & (0.06) & (0.07) \\
MIF & & & 0.47  &  0.51  &  0.55  & & &  0.47  &  0.49  &  0.53  & & &  0.52  &  0.53  &  0.55 \\
\setrow{\itshape} & & & (0.12) & (0.11) & (0.15) & & & (0.09) & (0.08) & (0.08) & & & (0.09) & (0.06) & (0.06) \\
\hline
$\gamma^*=0.4$ & & & & & & & & & & & & & & & \\
KM & & & 0.17  &  0.19  &  0.29  & & &  0.17  &  0.21  &  0.31  & & &  0.26  &  0.28  &  0.34 \\
\setrow{\itshape} & & & (0.19) & (0.09) & (0.09) & & & (0.06) & (0.08) & (0.08) & & & (0.07) & (0.06) & (0.08) \\
MIF & & & 0.20  &  0.21  &  0.29  & & &  0.22  &  0.24  &  0.31  & & &  0.31  &  0.31  &  0.35 \\
\setrow{\itshape} & & & (0.09) & (0.10) & (0.10) & & & (0.09) & (0.09) & (0.08) & & & (0.11) & (0.07) & (0.07) \\
\hline
$p^*=0.3$ & & & & & & & & & & & & & & & \\
KM & & & 0.08  &  0.11  &  0.19  & & &  0.08  &  0.12  &  0.21  & & &  0.16  &  0.17  &  0.24 \\
\setrow{\itshape} & & & (0.08) & (0.08) & (0.09) & & & (0.04) & (0.07) & (0.09) & & & (0.07) & (0.07) & (0.09) \\
MIF & & & 0.11  &  0.12  &  0.18  & & &  0.12  &  0.13  &  0.20  & & &  0.21  &  0.20  &  0.24 \\
\setrow{\itshape} & & & (0.10) & (0.09) & (0.08) & & & (0.09) & (0.07) & (0.08) & & & (0.12) & (0.07) & (0.07) \\
\hline
$i_0^*=0.01$ & & & & & & & & & & & & & & & \\
KM & & & 0.028  &  0.020  &  0.023  & & &  0.023  &  0.019  &  0.015  & & &  0.020  &  0.017  &  0.013 \\
\setrow{\itshape} & & & (0.069) & (0.022) & (0.078) & & & (0.016) & (0.012) & (0.010) & & & (0.009) & (0.006) & (0.006) \\
MIF & & & 0.025  &  0.022  &  0.023  & & &  0.022  &  0.020  &  0.015  & & &  0.018  &  0.015  &  0.013 \\
\setrow{\itshape} & & & (0.027) & (0.029) & (0.061) & & & (0.016) & (0.015) & (0.009) & & & (0.009) & (0.005) & (0.005) \\
\end{tabular}
\end{center}
\end{table}
}

\noindent The results on the second set of epidemic parameters displayed in Tables \ref{exp1_p08} and \ref{exp1_p03} are more contrasted, since the parameter values chosen ($\lambda^*=0.6$ and $\gamma^*=0.4$) generate more stochasticity (see Figure~\ref{exp1_obs}), so trajectories are less similar and further from the mean of the jump process; hence estimates are less accurate. Besides, the peak of the number of infectious individuals is clearly lower than in the $\lambda^*=1$ and $\gamma^*=1/3$ case. The estimates of $p$ are particularly poor when $n$ is low, which obviously impacts estimation of the other parameters. 

\subsubsection{Numerical results for the second experiment ($\tau \neq 0$)}

\paragraph{Unknown starting point $i_0$} Table \ref{exp1_p08_tau_s0_i0} displays the results obtained by our Kalman-based method and the MIF algorithm for the high-reporting scenario ($p=0.8$).

{\setlength{\tabcolsep}{2pt%red}
\begin{table}[H]
\begin{center}
\caption{Second experiment ($\tau \neq 0$). Estimation of $\theta=(\lambda,\gamma,p,i_0,\tau)$ under the constraint $s_0+i_0=1$ in Setting 1 with true parameter values $(\lambda^*,\gamma^*,p^*,i_0^*,\tau^*)$=$(0.6,0.4,0.8,0.01,0.5)$.
For each combination of $(N,n)$ and for each model parameter, point estimates and standard deviations are calculated as the mean of the $500$ individual estimates and their standard deviations (in brackets)
obtained by our Kalman-based method and the MIF algorithm. The reported values for the number of observations $n$ correspond to the average over the $500$ trajectories, with the min and max in brackets.} \label{exp1_p08_tau_s0_i0}
\footnotesize
\begin{tabular}{ccccccccccccccccc}
& & & &  \multicolumn{5}{c}{$N=1000$} & & & & \multicolumn{5}{c}{$N=10000$} \\
& & & & $n=10$ & $n=30$ & $n=99$ & $n=499$ & $n=998$ & & & & $n=11$ & $n=31$ & $n=101$ & $n=500$ & $n=1001$ \\
& & & & $(3,19)$ & $(9,56)$ & $(30,182)$ & $(152,916)$ & $(304,1831)$ & & & & $(8,17)$ & $(24,49)$ & $(78,160)$ & $(385,789)$ & $(771,1577)$ \\
\hline
$\lambda^*=0.6$ & & & & & & & & & & & & & & & & \\
KM & & & & 0.50 & 0.49 & 0.56 & 0.58 & 0.57 & & & & 0.48 & 0.50 & 0.55 & 0.57 & 0.58\\
\setrow{\itshape} & & & & (0.32) & (0.15) & (0.13) & (0.12) & (0.14) & & & & (0.04) & (0.07) & (0.06) & (0.05)& (0.07)\\
MIF & & & & 0.52 & 0.51 & 0.56 & 0.59 & 0.58 & & & & 0.52 & 0.52 & 0.56 & 0.58 & 0.59\\
\setrow{\itshape} & & & & (0.13) & (0.10) & (0.11) & (0.10) & (0.11) & & & & (0.04) & (0.05) & (0.05) & (0.04) & (0.05) \\
\hline
$\gamma^*=0.4$ & & & & & & & & & & & & & & & & \\
KM & & & & 0.20 & 0.25 & 0.35 & 0.39 & 0.39 & & & & 0.25 & 0.28 & 0.34 & 0.38 & 0.39\\
\setrow{\itshape} & & & & (0.33) & (0.13) & (0.09) & (0.07) & (0.09) & & & & (0.04) & (0.07) & (0.06) & (0.03) & (0.04)\\
MIF & & & & 0.25 & 0.30 & 0.36 & 0.39 & 0.38 & & & & 0.30 & 0.32 & 0.36 & 0.39 & 0.40\\
\setrow{\itshape} & & & & (0.07) & (0.07) & (0.08) & (0.07) & (0.07) & & & & (0.05) & (0.05) & (0.05) & (0.04) & (0.04) \\
\hline
$p^*=0.8$ & & & & & & & & & & & & & & & & \\
KM & & & & 0.24 & 0.42 & 0.66 & 0.77 & 0.78 & & & & 0.39 & 0.49 & 0.65 & 0.75 & 0.77\\
\setrow{\itshape} & & & & (0.16) & (0.25) & (0.23) & (0.16) & (0.14) & & & & (0.11) & (0.20) & (0.19) & (0.11) & (0.11)\\
MIF & & & & 0.39 & 0.50 & 0.65 & 0.72 & 0.72 & & & & 0.51 & 0.55 & 0.68 & 0.76 & 0.79\\
\setrow{\itshape} & & & & (0.16) & (0.17) & (0.17) & (0.13) & (0.15) & & & & (0.13) & (0.15) & (0.15) & (0.13) & (0.13) \\
\hline
$i_0^*=0.01$ & & & & & & & & & & & & & & & & \\
KM & & & & 0.029 & 0.037 & 0.022 & 0.016 & 0.015 & & & & 0.019 & 0.017 & 0.014 & 0.012 & 0.011\\
\setrow{\itshape} & & & & (0.048) & (0.086) & (0.078) & (0.039) & (0.010) & & & & (0.006) & (0.007) & (0.004) & (0.003) & (0.004)\\
MIF & & & & 0.014 & 0.016 & 0.014 & 0.012 & 0.012 & & & & 0.015 & 0.015 & 0.013 & 0.011 & 0.011\\
\setrow{\itshape} & & & & (0.005) & (0.005) & (0.004) & (0.003) & (0.003) & & & & (0.004) & (0.004) & (0.003) & (0.002) & (0.002) \\
\hline
$\tau^*=0.5$ & & & & & & & & & & & & & & & & \\
KM & & & & 0.34 & 0.44 & 0.52 & 0.54 & 0.52 & & & & 0.11 & 0.21 & 0.35 & 0.42 & 0.47\\
\setrow{\itshape} & & & & (0.69) & (0.44) & (0.20) & (0.16) & (0.17) & & & & (0.22) & (0.21) & (0.20) & (0.13) & (0.13)\\
MIF & & & & 0.49 & 0.49 & 0.47 & 0.48 & 0.43 & & & & 0.46 & 0.39 & 0.35 & 0.44 & 0.49\\
\setrow{\itshape} & & & & (0.25) & (0.19) & (0.17) & (0.15) & (0.20) & & & & (0.24) & (0.18) & (0.17) & (0.15) & (0.14) \\
\end{tabular}
\end{center}
\end{table}
}

\paragraph{Known starting point $i_0$} Tables \ref{exp1_p08_tau} and \ref{exp1_p03_tau} respectively display the results obtained by our Kalman-based method and the MIF algorithm for the high-reporting
scenario ($p=0.8$) and low-reporting scenario ($p=0.3$).

{\setlength{\tabcolsep}{3pt%red}
\begin{table}[H]
\begin{center}
\caption{Second experiment ($\tau \neq 0$). Estimation of $\theta=(\lambda,\gamma,p,\tau)$ with $s_0=0.99$ and $i_0=0.01$ known in Setting $1$ with true parameter values $(\lambda^*,\gamma^*,p^*,\tau^*)$=$(0.6,0.4,0.8,0.5)$.
For each combination of $(N,n)$ and for each model parameter, point estimates and standard deviations are calculated as the mean of the $500$ individual estimates and their standard deviations (in brackets)
obtained by KM and MIF. The reported values for the number of observations $n$ correspond to the average over the $500$ trajectories, with the min and max in brackets.} \label{exp1_p08_tau}
\footnotesize
\begin{tabular}{cccccccccccccccc}
& & &  \multicolumn{3}{c}{$N=1000$} & & & \multicolumn{3}{c}{$N=2000$} & & & \multicolumn{3}{c}{$N=10000$} \\
& & & $n=10$ & $n=30$ & $n=99$ & & & $n=11$ & $n=31$ & $n=102$ & & & $n=11$ & $n=31$ & $n=101$ \\
& & & $(3,19)$ & $(9,56)$ & $(30,182)$ & & & $(7,19)$ & $(20,56)$ & $(66,182)$ & & & $(8,17)$ & $(24,49)$ & $(78,160)$\\
\hline
$\lambda^*=0.6$ & & & & & & & & & & & & & & & \\
KM & & & 0.54 & 0.57 & 0.59 & & & 0.55 & 0.58 & 0.60 & & & 0.56 & 0.56 & 0.59\\
\setrow{\itshape} & & & (0.15) & (0.15) & (0.13) & & & (0.11) & (0.10) & (0.08) & & & (0.06) & (0.06) & (0.05) \\
MIF & & & 0.55 & 0.55 & 0.59 & & & 0.56 & 0.57 & 0.60 & & & 0.57 & 0.58 & 0.60\\
\setrow{\itshape} & & & (0.12) & (0.10) & (0.11) & & & (0.09) & (0.07) & (0.07) & & & (0.04) & (0.03) & (0.03) \\
\hline
$\gamma^*=0.4$ & & & & & & & & & & & & & & & \\
KM & & & 0.27 & 0.35 & 0.38 & & & 0.30 & 0.36 & 0.39 & & & 0.35 & 0.36 & 0.39\\
\setrow{\itshape} & & & (0.11) & (0.11) & (0.10) & & & (0.09) & (0.07) & (0.05) & & & (0.05) & (0.05) & (0.04) \\
MIF & & & 0.27 & 0.34 & 0.39 & & & 0.30 & 0.37 & 0.40 & & & 0.36 & 0.39 & 0.41\\
\setrow{\itshape} & & & (0.09) & (0.07) & (0.07) & & & (0.09) & (0.05) & (0.04) & & & (0.04) & (0.03) & (0.02) \\
\hline
$p^*=0.8$ & & & & & & & & & & & & & & & \\
KM & & & 0.42 & 0.61 & 0.72 & & & 0.49 & 0.67 & 0.76 & & & 0.65 & 0.69 & 0.77\\
\setrow{\itshape} & & & (0.22) & (0.22) & (0.18) & & & (0.20) & (0.19) & (0.15) & & & (0.14) & (0.15) & (0.12) \\
MIF & & & 0.43 & 0.62 & 0.75 & & & 0.49 & 0.67 & 0.78 & & & 0.67 & 0.75 & 0.80\\
\setrow{\itshape} & & & (0.19) & (0.17) & (0.13) & & & (0.19) & (0.14) & (0.10) & & & (0.12) & (0.09) & (0.07) \\
\hline
$\tau^*=0.5$ & & & & & & & & & & & & & & & \\
KM & & & 0.41 & 0.63 & 0.60 & & & 0.32 & 0.65 & 0.62 & & & 0.09 & 0.30 & 0.46\\
\setrow{\itshape} & & & (0.47) & (0.33) & (0.20) & & & (0.42) & (0.25) & (0.16) & & & (0.26) & (0.28) & (0.16) \\
MIF & & & 0.52 & 0.56 & 0.55 & & & 0.55 & 0.62 & 0.56 & & & 0.50 & 0.50 & 0.47\\
\setrow{\itshape} & & & (0.27) & (0.20) & (0.13) & & & (0.29) & (0.18) & (0.12) & & & (0.27) & (0.19) & (0.10) \\
\end{tabular}
\end{center}
\end{table}
}

{\setlength{\tabcolsep}{3pt%red}
\begin{table}[H]
\begin{center}
\caption{Second experiment ($\tau \neq 0$). Estimation of $\theta=(\lambda,\gamma,p,\tau)$ with $s_0=0.99$ and $i_0=0.01$ known in Setting $2$ with true parameter values $(\lambda^*,\gamma^*,p^*,\tau^*)$=$(0.6,0.4,0.3,0.5)$.
For each combination of $(N,n)$ and for each model parameter, point estimates and standard deviations are calculated as the mean of the $500$ individual estimates and their standard deviations (in brackets)
obtained by KM and MIF. The reported values for the number of observations $n$ correspond to the average over the $500$ trajectories, with the min and max in brackets.} \label{exp1_p03_tau}
\footnotesize
\begin{tabular}{cccccccccccccccc}
& & &  \multicolumn{3}{c}{$N=1000$} & & & \multicolumn{3}{c}{$N=2000$} & & &  \multicolumn{3}{c}{$N=10000$} \\
& & & $n=10$ & $n=30$ & $n=99$ & & & $n=11$ & $n=31$ & $n=102$ & & & $n=11$ & $n=31$ & $n=101$ \\
& & & $(3,19)$ & $(9,56)$ & $(30,182)$ & & & $(7,19)$ & $(20,56)$ & $(66,182)$ & & & $(8,17)$ & $(24,49)$ & $(78,160)$\\
\hline
$\lambda^*=0.6$ & & & & & & & & & & & & & & & \\
KM & & & 0.47 & 0.57 & 0.57 & & & 0.47 & 0.54 & 0.58 & & & 0.56 & 0.55 & 0.56\\
\setrow{\itshape} & & & (0.16) & (0.13) & (0.16) & & & (0.12) & (0.09) & (0.09) & & & (0.06) & (0.06) & (0.05) \\
MIF & & & 0.51 & 0.55 & 0.55 & & & 0.52 & 0.55 & 0.56 & & & 0.57 & 0.57 & 0.59\\
\setrow{\itshape} & & & (0.12) & (0.13) & (0.11) & & & (0.09) & (0.07) & (0.07) & & & (0.05) & (0.04) & (0.03) \\
\hline
$\gamma^*=0.4$ & & & & & & & & & & & & & & & \\
KM & & & 0.20 & 0.29 & 0.35 & & & 0.26 & 0.30 & 0.37 & & & 0.37 & 0.35 & 0.35\\
\setrow{\itshape} & & & (0.12) & (0.13) & (0.14) & & & (0.10) & (0.09) & (0.09) & & & (0.05) & (0.06) & (0.05) \\
MIF & & & 0.22 & 0.26 & 0.32 & & & 0.25 & 0.30 & 0.34 & & & 0.38 & 0.37 & 0.39\\
\setrow{\itshape} & & & (0.08) & (0.10) & (0.09) & & & (0.10) & (0.08) & (0.06) & & & (0.06) & (0.04) & (0.03) \\
\hline
$p^*=0.3$ & & & & & & & & & & & & & & & \\
KM & & & 0.10 & 0.19 & 0.24 & & & 0.15 & 0.20 & 0.27 & & & 0.28 & 0.25 & 0.25\\
\setrow{\itshape} & & & (0.09) & (0.15) & (0.12) & & & (0.08) & (0.12) & (0.13) & & & (0.06) & (0.08) & (0.10) \\
MIF & & & 0.11 & 0.14 & 0.19 & & & 0.14 & 0.18 & 0.21 & & & 0.27 & 0.27 & 0.28\\
\setrow{\itshape} & & & (0.06) & (0.09) & (0.06) & & & (0.08) & (0.07) & (0.05) & & & (0.07) & (0.04) & (0.03) \\
\hline
$\tau^*=0.5$ & & & & & & & & & & & & & & & \\
KM & & & 0.23 & 0.42 & 0.52 & & & 0.13 & 0.24 & 0.51 & & & 0.05 & 0.19 & 0.36\\
\setrow{\itshape} & & & (0.20) & (0.29) & (0.21) & & & (0.19) & (0.21) & (0.18) & & & (0.14) & (0.15) & (0.12) \\
MIF & & & 0.38 & 0.37 & 0.44 & & & 0.39 & 0.29 & 0.43 & & & 0.61 & 0.33 & 0.43\\
\setrow{\itshape} & & & (0.14) & (0.15) & (0.09) & & & (0.15) & (0.15) & (0.07) & & & (0.17) & (0.12) & (0.05) \\
\end{tabular}
\end{center}
\end{table}
}

\subsection{Numerical confidence intervals}

 Figures \ref{exp1_lambda_profile} and \ref{exp1_gamma_profile} represent the profile likelihoods and the subsequent confidence intervals (CI$95\%$) for the parameters $\lambda$ and $\gamma$ obtained for our Kalman filtering-based method in two settings (first case:  $N=2000$, $n=30$, and $p=0.3$; second case: $N=10000$, $n=100$, and $p=0.8$).
\begin{center}
   \begin{minipage}[b]{0.48\linewidth}
      \centering \includegraphics[scale=0.16]{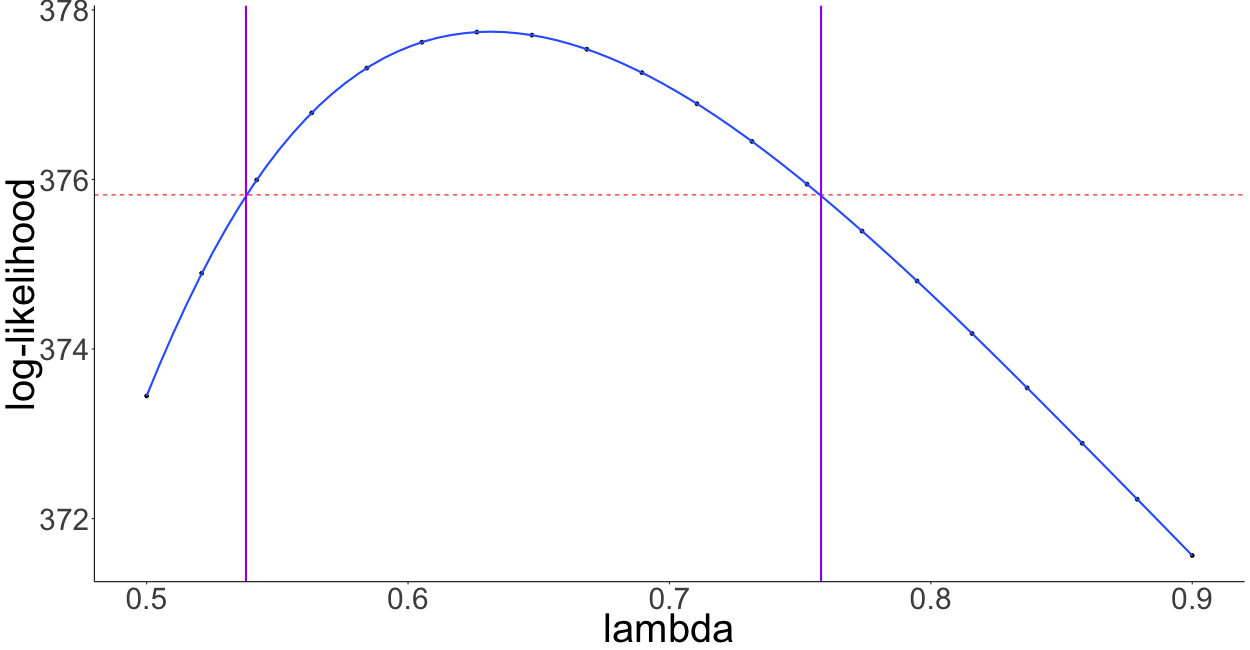}
   \end{minipage}\hfill
   \begin{minipage}[b]{0.48\linewidth}   
      \centering \includegraphics[scale=0.16]{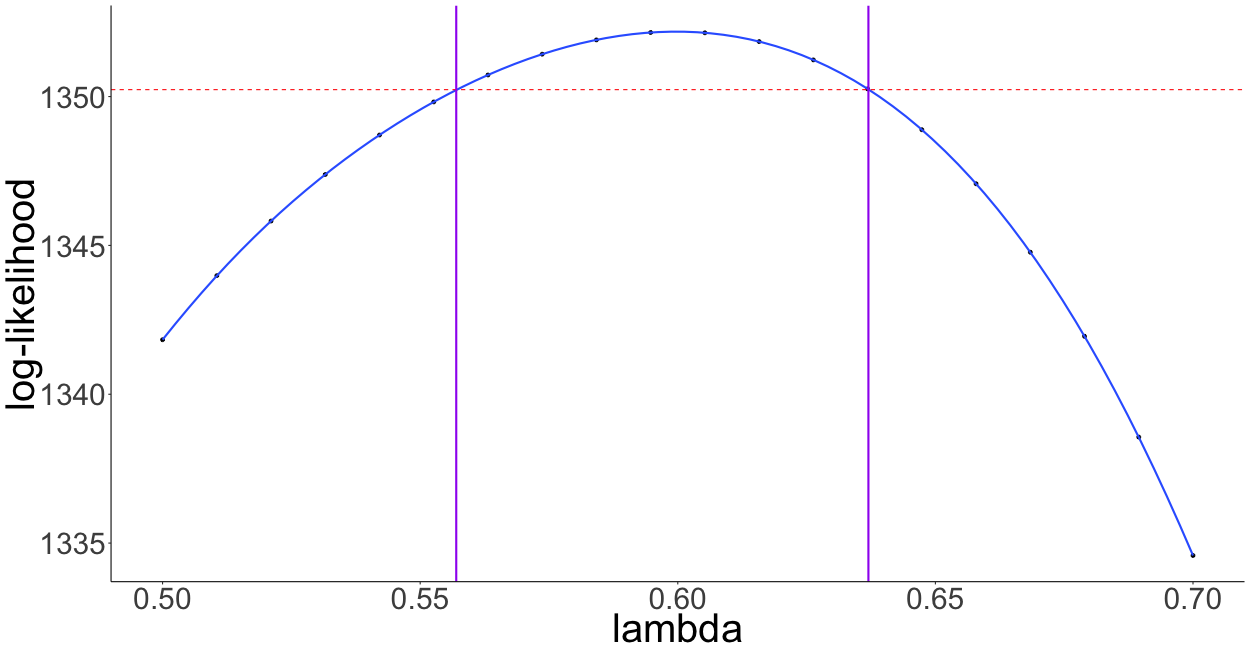}
   \end{minipage}\hfill
\begin{figure}[H]
 \caption{Profile likelihood and confidence intervals (CI$95\%$) for $\lambda$. Left panel: $N=2000$, $n=30$, and $p=0.3$. The true value $\lambda^*=0.6$, the point estimate $\hat{\lambda}=0.47$, and CI$95\% = [0.54,0.76]$. Right panel: $N=10000$, $n=100$, and $p=0.8$. The true value $\lambda^*=0.6$, the point estimate $\hat{\lambda}=0.60$, and CI$95\% = [0.56,0.64]$.} \label{exp1_lambda_profile}
\end{figure}
\end{center} 
\begin{center}
   \begin{minipage}[b]{0.48\linewidth}
      \centering \includegraphics[scale=0.16]{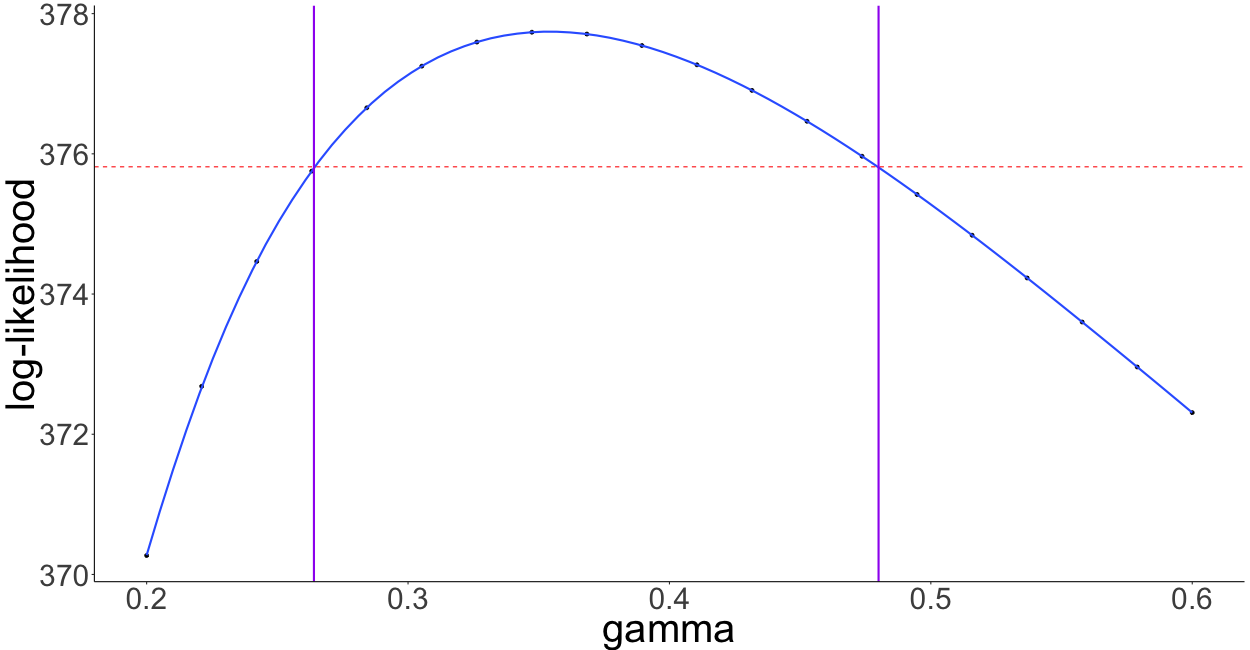}
   \end{minipage}\hfill
   \begin{minipage}[b]{0.48\linewidth}   
      \centering \includegraphics[scale=0.16]{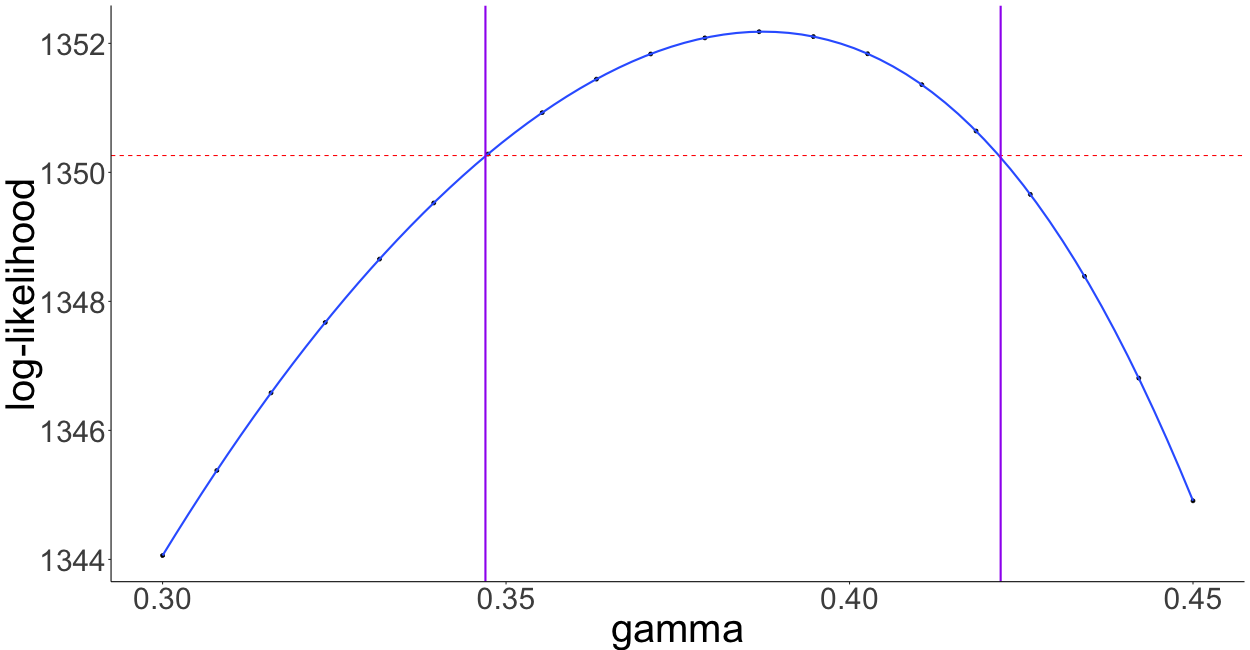}
   \end{minipage}\hfill
\begin{figure}[H]
 \caption{Profile likelihood and confidence intervals (CI$95\%$) for $\gamma$. Left panel: $N=2000$, $n=30$, and $p=0.3$. The true value $\gamma^*=0.4$, the point estimate $\hat{\gamma}=0.21$, and CI$95\% = [0.26,0.48]$. Right panel: $N=10000$, $n=100$, and $p=0.8$. The true value $\gamma^*=0.4$, the point estimate $\hat{\gamma}=0.40$, and CI$95\% = [0.35,0.42]$.} \label{exp1_gamma_profile}
\end{figure}
\end{center} 

\section{User-friendly code}
\label{sec:codeR}

We propose user-friendly code composed of four distinct programs in the R language, available at the
 RunMyCode website: \url{http://www.runmycode.org/companion/view/4074}.

\begin{itemize}

\item \textit{KalmanFunctions.R} includes general functions implementing the Kalman filter and computing the likelihood of the observations, given a specified compartmental
model, with a fixed sampling interval. These functions are easily generalizable to the case where the sampling interval is variable. Moreover, this script includes a function computing the
resolvent matrix defined in (\ref{dPhi}) for large time intervals between observations $\Delta$. 

\item \textit{ModelFunctions.R} implements the SIR and SEIR models and defines the key quantities (described in the manuscript for the SIR model) necessary to apply the Kalman filter-based method. More precisely, given a compartmental model (SIR or SEIR), the following functions are implemented: the ode system, the drift function, the gradient of the drift function, the diffusion 
matrix, the projection operator linking the observations to the states of the epidemic model and the variance of the observations.

\item \textit{SIRexample.R} and \textit{SEIRexample.R} simulate
respectively SIR and SEIR Markovian jump processes for 
a set of parameters values, using the \texttt{GillespieSSA} package. The observations of infectious individuals are obtained
by: $O_1(t_k)\sim \text{Binomial}(I(t_k), p), \quad O_2(t_k)\sim {\cal N}(0,\tau^2 I(t_k))$, $k=1,\ldots,n$, at regularly-spaced time points. Finally, an estimation of key parameters
$\lambda$, $\gamma$, $p$ and $\tau$ with known starting points and, in the SEIR model, with a known transition rate from E to I, is proposed. 

\end{itemize}

\bibliographystyle{elsarticle-harv}

\end{document}